\pdfoutput=1
\documentclass[12pt,a4paper]{article}

\usepackage{ifthen}
\newboolean{pdflatex}
\setboolean{pdflatex}{true}

\newboolean{articletitles}
\setboolean{articletitles}{true}

\newboolean{uprightparticles}
\setboolean{uprightparticles}{false}

\newboolean{inbibliography}
\setboolean{inbibliography}{false}

\def\paperauthors{LHCb collaboration}
\def\paperasciititle{Amplitude analysis of Bs0 -> KS0 K+- pi-+ decays}
\def\papertitle{Amplitude analysis of $\Bs \to \KS\Kpm\pimp$ decays}
\def\paperkeywords{{High Energy Physics}, {LHCb}}
\def\papercopyright{\the\year\ CERN for the benefit of the LHCb collaboration}
\def\paperlicence{CC-BY-4.0 licence}
\def\paperlicenceurl{https://creativecommons.org/licenses/by/4.0/}

\usepackage{multirow}
\usepackage{arydshln}

\usepackage[top=1in, bottom=1.25in, left=1in, right=1in]{geometry}

\usepackage{microtype}
\usepackage{lineno}  
\usepackage{xspace} 
\usepackage{caption} 

\usepackage{graphicx}  
\usepackage{color}
\usepackage{colortbl}
\graphicspath{{./figs/}} 
\DeclareGraphicsExtensions{.pdf,.PDF,png,.PNG}

\usepackage{amsmath} 
\usepackage{amssymb}
\usepackage{amsfonts}
\usepackage{upgreek} 

\newcommand*\patchAmsMathEnvironmentForLineno[1]{%
\expandafter\let\csname old#1\expandafter\endcsname\csname #1\endcsname
\expandafter\let\csname oldend#1\expandafter\endcsname\csname
end#1\endcsname
 \renewenvironment{#1}%
   {\linenomath\csname old#1\endcsname}%
   {\csname oldend#1\endcsname\endlinenomath}%
}
\newcommand*\patchBothAmsMathEnvironmentsForLineno[1]{%
  \patchAmsMathEnvironmentForLineno{#1}%
  \patchAmsMathEnvironmentForLineno{#1*}%
}
\AtBeginDocument{%
\patchBothAmsMathEnvironmentsForLineno{equation}%
\patchBothAmsMathEnvironmentsForLineno{align}%
\patchBothAmsMathEnvironmentsForLineno{flalign}%
\patchBothAmsMathEnvironmentsForLineno{alignat}%
\patchBothAmsMathEnvironmentsForLineno{gather}%
\patchBothAmsMathEnvironmentsForLineno{multline}%
\patchBothAmsMathEnvironmentsForLineno{eqnarray}%
}

\usepackage{hyperxmp}

\usepackage[pdftex,
            pdfauthor={\paperauthors},
            pdftitle={\paperasciititle},
            pdfkeywords={\paperkeywords},
            pdfcopyright={Copyright (C) \papercopyright},
            pdflicenseurl={\paperlicenceurl}]{hyperref}

\usepackage[colorinlistoftodos,textsize=scriptsize]{todonotes}

\usepackage[all]{hypcap} 

\usepackage{xspace} 
\usepackage{upgreek}

\def\lhcb {\mbox{LHCb}\xspace}

\def\velo   {VELO\xspace}

\def\MagUp {\mbox{\em Mag\kern -0.05em Up}\xspace}

\ifthenelse{\boolean{uprightparticles}}%
{

 \def\Ppi         {\ensuremath{\uppi}\xspace}

 \def\Ppsi        {\ensuremath{\uppsi}\xspace}

 \def\PDelta      {\ensuremath{\Delta}\xspace}                 
 \def\PXi      {\ensuremath{\Xi}\xspace}                 
 \def\PLambda      {\ensuremath{\Lambda}\xspace}                 
 \def\PSigma      {\ensuremath{\Sigma}\xspace}                 
 \def\POmega      {\ensuremath{\Omega}\xspace}                 
 \def\PUpsilon      {\ensuremath{\Upsilon}\xspace}                 
 
 \def\PB      {\ensuremath{\mathrm{B}}\xspace}                 
                  
 \def\PD      {\ensuremath{\mathrm{D}}\xspace}

 \def\PJ      {\ensuremath{\mathrm{J}}\xspace}                 
 \def\PK      {\ensuremath{\mathrm{K}}\xspace}

 \def\Pb      {\ensuremath{\mathrm{b}}\xspace}                 
 \def\Pc      {\ensuremath{\mathrm{c}}\xspace}

 \def\Ph      {\ensuremath{\mathrm{h}}\xspace}                 
 \def\Pi      {\ensuremath{\mathrm{i}}\xspace}

 \def\Pp      {\ensuremath{\mathrm{p}}\xspace}

 \def\Ps      {\ensuremath{\mathrm{s}}\xspace}

}
{

 \def\Ppi         {\ensuremath{\pi}\xspace}

 \def\Ppsi        {\ensuremath{\psi}\xspace}                 
                  
 \mathchardef\PDelta="7101
 \mathchardef\PXi="7104
 \mathchardef\PLambda="7103
 \mathchardef\PSigma="7106
 \mathchardef\POmega="710A
 \mathchardef\PUpsilon="7107
                  
 \def\PB      {\ensuremath{B}\xspace}                 
                  
 \def\PD      {\ensuremath{D}\xspace}

 \def\PJ      {\ensuremath{J}\xspace}                 
 \def\PK      {\ensuremath{K}\xspace}

 \def\Pb      {\ensuremath{b}\xspace}                 
 \def\Pc      {\ensuremath{c}\xspace}

 \def\Ph      {\ensuremath{h}\xspace}                 
 \def\Pi      {\ensuremath{i}\xspace}

 \def\Pp      {\ensuremath{p}\xspace}

 \def\Ps      {\ensuremath{s}\xspace}

}

\makeatletter
\ifcase \@ptsize \relax
  \newcommand{\miniscule}{\@setfontsize\miniscule{4}{5}}
\or
  \newcommand{\miniscule}{\@setfontsize\miniscule{5}{6}}
\or
  \newcommand{\miniscule}{\@setfontsize\miniscule{5}{6}}
\fi
\makeatother

\DeclareRobustCommand{\optbar}[1]{\shortstack{{\miniscule (\rule[.5ex]{1.25em}{.18mm})}
  \\ [-.7ex] $#1$}}

\def\squark    {{\ensuremath{\Ps}}\xspace}

\def\cquark    {{\ensuremath{\Pc}}\xspace}
\def\cquarkbar {{\ensuremath{\overline \cquark}}\xspace}

\def\bquark    {{\ensuremath{\Pb}}\xspace}

\def\pion   {{\ensuremath{\Ppi}}\xspace}

\def\pip    {{\ensuremath{\pion^+}}\xspace}
\def\pim    {{\ensuremath{\pion^-}}\xspace}
\def\pipm   {{\ensuremath{\pion^\pm}}\xspace}
\def\pimp   {{\ensuremath{\pion^\mp}}\xspace}

\def\kaon    {{\ensuremath{\PK}}\xspace}
  \def\Kbar    {{\kern 0.2em\overline{\kern -0.2em \PK}{}}\xspace}

\def\KorKbar    {\kern 0.18em\optbar{\kern -0.18em K}{}\xspace}
\def\Kz      {{\ensuremath{\kaon^0}}\xspace}
\def\Kzb     {{\ensuremath{\Kbar{}^0}}\xspace}
\def\Kp      {{\ensuremath{\kaon^+}}\xspace}
\def\Km      {{\ensuremath{\kaon^-}}\xspace}
\def\Kpm     {{\ensuremath{\kaon^\pm}}\xspace}
\def\Kmp     {{\ensuremath{\kaon^\mp}}\xspace}
\def\KS      {{\ensuremath{\kaon^0_{\mathrm{ \scriptscriptstyle S}}}}\xspace}

\def\Kstarz  {{\ensuremath{\kaon^{*0}}}\xspace}
\def\Kstarzb {{\ensuremath{\Kbar{}^{*0}}}\xspace}
\def\Kstar   {{\ensuremath{\kaon^*}}\xspace}
\def\Kstarb  {{\ensuremath{\Kbar{}^*}}\xspace}
\def\Kstarp  {{\ensuremath{\kaon^{*+}}}\xspace}
\def\Kstarm  {{\ensuremath{\kaon^{*-}}}\xspace}
\def\Kstarpm {{\ensuremath{\kaon^{*\pm}}}\xspace}

\def\KorKbarz {\ensuremath{\KorKbar^0}\xspace}

\def\Dbar    {{\kern 0.2em\overline{\kern -0.2em \PD}{}}\xspace}
\def\D       {{\ensuremath{\PD}}\xspace}

\def\DorDbar    {\kern 0.18em\optbar{\kern -0.18em D}{}\xspace}
\def\Dz      {{\ensuremath{\D^0}}\xspace}

\def\Dstarp  {{\ensuremath{\D^{*+}}}\xspace}

\def\DorDsm     {{\ensuremath{\D^-_{(\squark)}}}\xspace}

\def\B       {{\ensuremath{\PB}}\xspace}
\def\Bbar    {{\ensuremath{\kern 0.18em\overline{\kern -0.18em \PB}{}}}\xspace}

\def\BorBbar    {\kern 0.18em\optbar{\kern -0.18em B}{}\xspace}
\def\Bz      {{\ensuremath{\B^0}}\xspace}

\def\Bu      {{\ensuremath{\B^+}}\xspace}

\def\Bp      {{\ensuremath{\Bu}}\xspace}

\def\Bs      {{\ensuremath{\B^0_\squark}}\xspace}
\def\Bsb     {{\ensuremath{\Bbar{}^0_\squark}}\xspace}

\def\BdorBs  {{\ensuremath{\B^0_{(\squark)}}}\xspace}

\def\jpsi     {{\ensuremath{{\PJ\mskip -3mu/\mskip -2mu\Ppsi\mskip 2mu}}}\xspace}

  \def\Y#1S{\ensuremath{\PUpsilon{(#1S)}}\xspace}

\def\proton      {{\ensuremath{\Pp}}\xspace}
\def\antiproton  {{\ensuremath{\overline \proton}}\xspace}

\def\Lz          {{\ensuremath{\PLambda}}\xspace}
\def\Lbar        {{\ensuremath{\kern 0.1em\overline{\kern -0.1em\PLambda}}}\xspace}
\def\LorLbar    {\kern 0.18em\optbar{\kern -0.18em \PLambda}{}\xspace}

\def\Lb      {{\ensuremath{\Lz^0_\bquark}}\xspace}
\def\Lbbar   {{\ensuremath{\Lbar{}^0_\bquark}}\xspace}

\def\Lcbar   {{\ensuremath{\Lbar{}^-_\cquark}}\xspace}

\def\BF         {{\ensuremath{\mathcal{B}}}\xspace}

\newcommand{\decay}[2]{\ensuremath{#1\!\to #2}\xspace}         

\def\to                 {\ensuremath{\rightarrow}\xspace}

\def\CP                {{\ensuremath{C\!P}}\xspace}

\def\AT#1     {\ensuremath{A_{\mathrm{T}}^{#1}}\xspace}           

\def\C#1      {\ensuremath{\mathcal{C}_{#1}}\xspace}                       
\def\Cp#1     {\ensuremath{\mathcal{C}_{#1}^{'}}\xspace}                    
\def\Ceff#1   {\ensuremath{\mathcal{C}_{#1}^{\mathrm{(eff)}}}\xspace}        
\def\Cpeff#1  {\ensuremath{\mathcal{C}_{#1}^{'\mathrm{(eff)}}}\xspace}       
\def\Ope#1    {\ensuremath{\mathcal{O}_{#1}}\xspace}                       
\def\Opep#1   {\ensuremath{\mathcal{O}_{#1}^{'}}\xspace}                    

\newcommand{\tev}{\ifthenelse{\boolean{inbibliography}}{\ensuremath{~T\kern -0.05em eV}}{\ensuremath{\mathrm{\,Te\kern -0.1em V}}}\xspace}
\newcommand{\gev}{\ensuremath{\mathrm{\,Ge\kern -0.1em V}}\xspace}
\newcommand{\mev}{\ensuremath{\mathrm{\,Me\kern -0.1em V}}\xspace}
\newcommand{\kev}{\ensuremath{\mathrm{\,ke\kern -0.1em V}}\xspace}
\newcommand{\ev}{\ensuremath{\mathrm{\,e\kern -0.1em V}}\xspace}
\newcommand{\gevc}{\ensuremath{{\mathrm{\,Ge\kern -0.1em V\!/}c}}\xspace}
\newcommand{\mevc}{\ensuremath{{\mathrm{\,Me\kern -0.1em V\!/}c}}\xspace}
\newcommand{\gevcc}{\ensuremath{{\mathrm{\,Ge\kern -0.1em V\!/}c^2}}\xspace}
\newcommand{\gevgevcccc}{\ensuremath{{\mathrm{\,Ge\kern -0.1em V^2\!/}c^4}}\xspace}
\newcommand{\mevcc}{\ensuremath{{\mathrm{\,Me\kern -0.1em V\!/}c^2}}\xspace}

\def\mum  {\ensuremath{{\,\upmu\mathrm{m}}}\xspace}

\def\invfb   {\ensuremath{\mbox{\,fb}^{-1}}\xspace}

\newcommand{\chisq}{\ensuremath{\chi^2}\xspace}

\def\gsim{{~\raise.15em\hbox{$>$}\kern-.85em
          \lower.35em\hbox{$\sim$}~}\xspace}
\def\lsim{{~\raise.15em\hbox{$<$}\kern-.85em
          \lower.35em\hbox{$\sim$}~}\xspace}

\def\pt         {\ensuremath{p_{\mathrm{ T}}}\xspace}
\def\ptot       {\ensuremath{p}\xspace}

\def\evtgen     {\mbox{\textsc{EvtGen}}\xspace}

\def\geant      {\mbox{\textsc{Geant4}}\xspace}

\def\photos     {\mbox{\textsc{Photos}}\xspace}

\def\pythia     {\mbox{\textsc{Pythia}}\xspace}

\def\tell1  {TELL1\xspace}
\def\ukl1   {UKL1\xspace}

\newcommand{\eg}{\mbox{\itshape e.g.}\xspace}
\newcommand{\ie}{\mbox{\itshape i.e.}\xspace}

\usepackage{multirow}
\usepackage{arydshln}

\usepackage{cite} 
\usepackage{mciteplus}

\allowdisplaybreaks

\ifthenelse{\boolean{uprightparticles}}%
{
 
}
{
 
}

\def\had  {\ensuremath{\Ph}\xspace}
\def\hadp  {\ensuremath{\Ph^+}\xspace}

\def\hadprimm {\ensuremath{\had^{\prime-}}\xspace}

\def\BdorBdbar    {\ensuremath{\kern 0.18em\optbar{\kern -0.18em B}{}^0}\xspace}
\def\BsorBsbar    {\ensuremath{\kern 0.18em\optbar{\kern  0.06em B_s}{}^0}\xspace}

\def\kpi       {\ensuremath{\kaon\pion}\xspace}

\def\KstarI        {\ensuremath{\Kstar(892)}\xspace}
\def\KstarIp       {\ensuremath{\Kstar(892)^+}\xspace}
\def\KstarIm       {\ensuremath{\Kstar(892)^-}\xspace}
\def\KstarIpm      {\ensuremath{\Kstar(892)^{\pm}}\xspace}
\def\KstarIz       {\ensuremath{\Kstar(892)^0}\xspace}
\def\KstarIzb      {\ensuremath{\Kstarb(892)^0}\xspace}
\def\KstarIzoptbar {\ensuremath{\KorKbar\!^*(892)^0}\xspace}

\def\KstarII        {\ensuremath{\kaon^*_0(1430)}\xspace}

\def\KstarIIp       {\ensuremath{\kaon^*_0(1430)^+}\xspace}
\def\KstarIIm       {\ensuremath{\kaon^*_0(1430)^-}\xspace}
\def\KstarIIpm      {\ensuremath{\kaon^*_0(1430)^{\pm}}\xspace}
\def\KstarIIz       {\ensuremath{\kaon^*_0(1430)^0}\xspace}
\def\KstarIIzb      {\ensuremath{\Kbar{}^*_0(1430)^0}\xspace}
\def\KstarIIzoptbar {\ensuremath{\KorKbar\!^*_0(1430)^0}\xspace}

\def\KstarIII        {\ensuremath{\kaon^*_2(1430)}\xspace}
\def\KstarIIIp       {\ensuremath{\kaon^*_2(1430)^+}\xspace}
\def\KstarIIIm       {\ensuremath{\kaon^*_2(1430)^-}\xspace}
\def\KstarIIIpm      {\ensuremath{\kaon^*_2(1430)^{\pm}}\xspace}
\def\KstarIIIz       {\ensuremath{\kaon^*_2(1430)^0}\xspace}
\def\KstarIIIzb      {\ensuremath{\Kbar{}^*_2(1430)^0}\xspace}
\def\KstarIIIzoptbar {\ensuremath{\KorKbar\!^*_2(1430)^0}\xspace}

\def\KpiS   {\ensuremath{(\kaon\pion)^*_0}\xspace}
\def\KpiSz  {\ensuremath{(\Kmp\pipm)^*_0}\xspace}
\def\KpiSpm {\ensuremath{(\KorKbarz\pipm)^*_0}\xspace}

\def\KpiNRz  {\ensuremath{(\Kmp\pipm)_\mathrm{NR}}\xspace}
\def\KpiNRpm {\ensuremath{(\KorKbarz\pipm)_\mathrm{NR}}\xspace}

\def\BstoKsKPi   {\decay{\Bs}{\KS \Kpm \pimp}}

\def\Kshhp{\ensuremath{\KS \hadp \hadprimm}\xspace}

\def\KsPiPi{\ensuremath{\KS \pip \pim}\xspace}

\def\KsKK{\ensuremath{\KS \Kp \Km}\xspace}

\def\KsKpPim{\ensuremath{\KS \Kp \pim}\xspace}
\def\KsKmPip{\ensuremath{\KS \Km \pip}\xspace}

\def\BdstoKshhp   {\decay{\BdorBs}{\KS \hadp \hadprimm}}

\def\LbtoKsPip         {\decay{\Lb}{\KS \proton \pim}}

\def\LL   {long\xspace}
\def\DD   {downstream\xspace}

\newcommand{\Br}[1]{\ensuremath{\BF\left(#1\right)}\xspace}

\newcommand{\tab}[1]{Table~\ref{tab : #1}}

\newcommand{\fig}[1]{Fig.~\ref{fig : #1}}

\def\phz {\phantom{0}}
\def\pho {\phantom{1}}

\newcommand{\mrthree}[1]{\multirow{3}{*}{#1}}
\newcommand{\mrsix}[1]{\multirow{6}{*}{#1}}

\def\BsToKzBarOptKpi{\decay{\Bs}{\KorKbar^{0} \Kpm \pimp}}

\def\KstarzorKstarzb {{\ensuremath{\KorKbar\!^{*0}}}\xspace}

\begin{document}

\renewcommand{\thefootnote}{\fnsymbol{footnote}}
\setcounter{footnote}{1}


\begin{titlepage}
\pagenumbering{roman}

\vspace*{-1.5cm}
\centerline{\large EUROPEAN ORGANIZATION FOR NUCLEAR RESEARCH (CERN)}
\vspace*{1.5cm}
\noindent
\begin{tabular*}{\linewidth}{lc@{\extracolsep{\fill}}r@{\extracolsep{0pt}}}
\ifthenelse{\boolean{pdflatex}}
{\vspace*{-1.5cm}\mbox{\!\!\!\includegraphics[width=.14\textwidth]{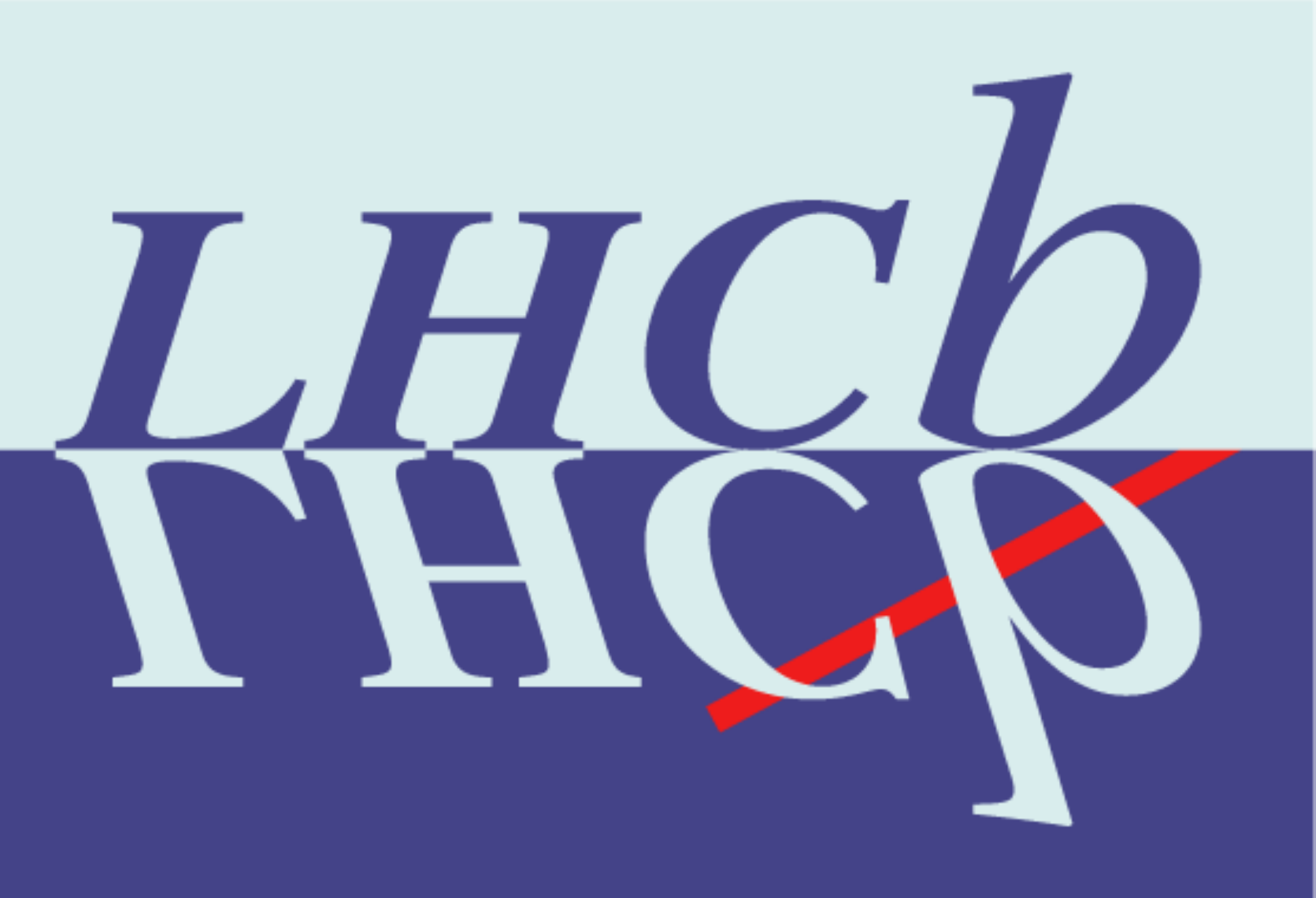}} & &}%
{\vspace*{-1.2cm}\mbox{\!\!\!\includegraphics[width=.12\textwidth]{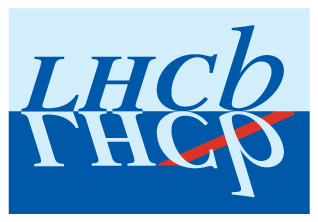}} & &}%
\\
 & & CERN-EP-2019-017 \\  
 & & LHCb-PAPER-2018-045 \\  
 & & \today \\ 
 & & \\
\end{tabular*}

\vspace*{4.0cm}

{\normalfont\bfseries\boldmath\huge
\begin{center}
  \papertitle 
\end{center}
}

\vspace*{2.0cm}

\begin{center}
\paperauthors\footnote{Authors are listed at the end of this paper.}
\end{center}

\vspace{\fill}

\begin{abstract}
  \noindent
  The first untagged decay-time-integrated amplitude analysis of
  $\Bs \to \KS\Kpm\pimp$ decays is performed using a sample
  corresponding to $3.0\invfb$ of $pp$ collision data recorded with the LHCb detector during 2011 and 2012.
  The data are described with an amplitude model that contains contributions
  from the intermediate resonances $\Kstar(892)^{0,+}$, $K^*_2(1430)^{0,+}$ and
  $K^*_0(1430)^{0,+}$, and their charge conjugates.
  Measurements of the branching fractions of the decay modes $\Bs\to K^{*}(892)^{\pm}\Kmp$ and $\Bs\to \KorKbar\!^{*}(892)^{0}\KorKbar^{0}$ are in agreement with, and more precise than, previous results. 
  The decays $\Bs\to \KstarIIpm\Kmp$ and
  $\Bs \to \KorKbar\!^{*}_{0}(1430)^0\KorKbar^{0}$ are observed for the
  first time, each with significance over 10 standard deviations.
\end{abstract}

\vspace*{2.0cm}

\begin{center}
  Published in JHEP 06 (2019) 114
\end{center}

\vspace{\fill}

{\footnotesize 
\centerline{\copyright~\papercopyright. \href{\paperlicenceurl}{\paperlicence}.}}
\vspace*{2mm}

\end{titlepage}


\newpage
\setcounter{page}{2}
\mbox{~}

\cleardoublepage

\renewcommand{\thefootnote}{\arabic{footnote}}
\setcounter{footnote}{0}


\pagestyle{plain} 
\setcounter{page}{1}
\pagenumbering{arabic}
\section{Introduction}
\label{sec:Introduction}

The search for new sources of \CP violation in addition to that predicted by 
the CKM matrix~\cite{Cabibbo:1963yz,Kobayashi:1973fv} is among the main goals of particle physics research.
One interesting approach is the study of decay-time distributions of neutral \B-meson decays to hadronic final states mediated by loop (\emph{penguin}) $b \to s$ amplitudes.  
As-yet undiscovered particles could contribute in the loops and cause the observables to deviate from the values expected in the Standard Model (SM)~\cite{Grossman:1996ke,Fleischer:1996bv,London:1997zk,Ciuchini:1997zp}.
Studies of various \Bz decays have been performed for this reason, including decay-time-dependent amplitude analyses of $B^0 \to \KS\pi^+\pi^-$~\cite{Dalseno:2008wwa,Aubert:2009me} and $B^0 \to \KS K^+K^-$~\cite{Nakahama:2010nj,Lees:2012kxa} transitions.
Such analyses, which involve describing the variation of the decay amplitudes over the phase-space of the three-body decays, are more sensitive to interference effects than the \emph{quasi-two-body} approach and are therefore particularly important when broad resonances contribute.
Decay-time-dependent analyses of \Bs-meson transitions mediated by hadronic $b \to s$ amplitudes have been performed for the $\Bs \to \Kp\Km$~\cite{LHCb-PAPER-2018-006}, $\Bs\to\phi\phi$~\cite{LHCb-PAPER-2014-026,LHCb-CONF-2018-001} and $\Bs\to\Kstarz\Kstarzb$~\cite{LHCb-PAPER-2017-048} decays, but not yet for any three-body \Bs\ decay.

\begin{figure}[!tb]
  \centering
  \includegraphics*[width=0.46\textwidth]{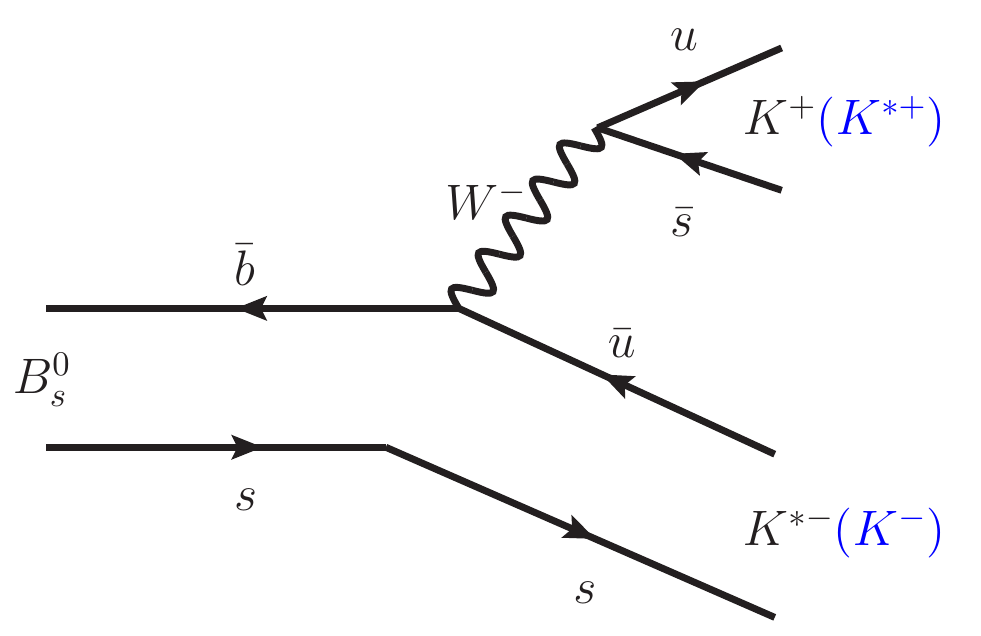}
  \includegraphics*[width=0.46\textwidth]{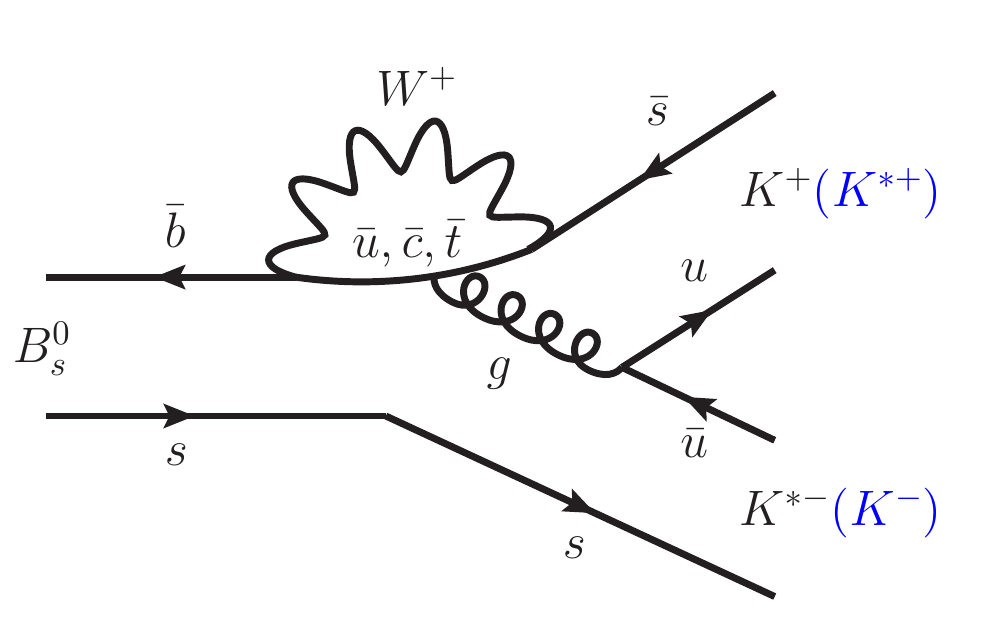}
  \includegraphics*[width=0.46\textwidth]{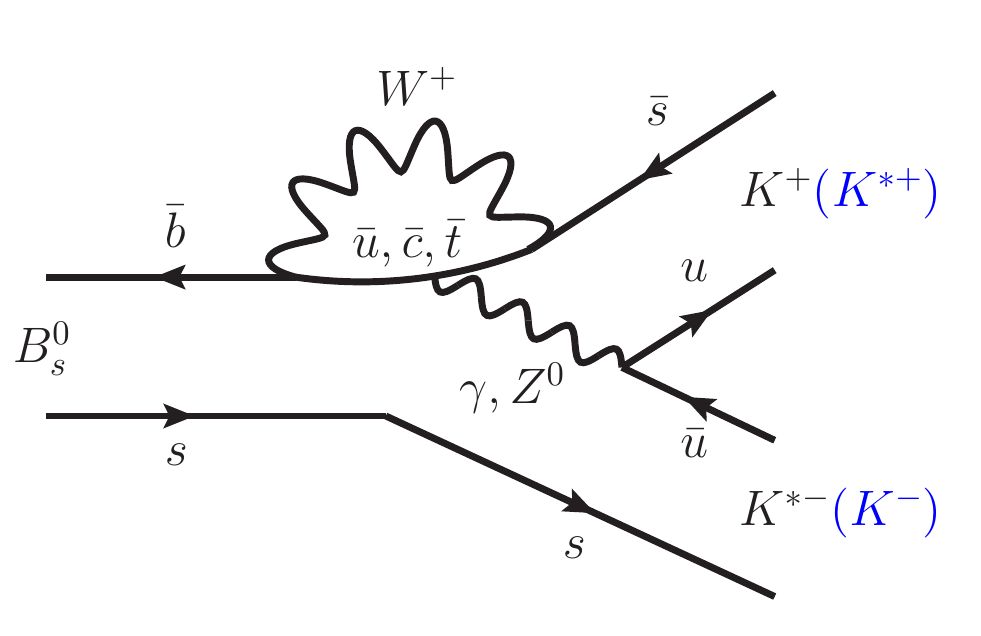}
  \includegraphics*[width=0.46\textwidth]{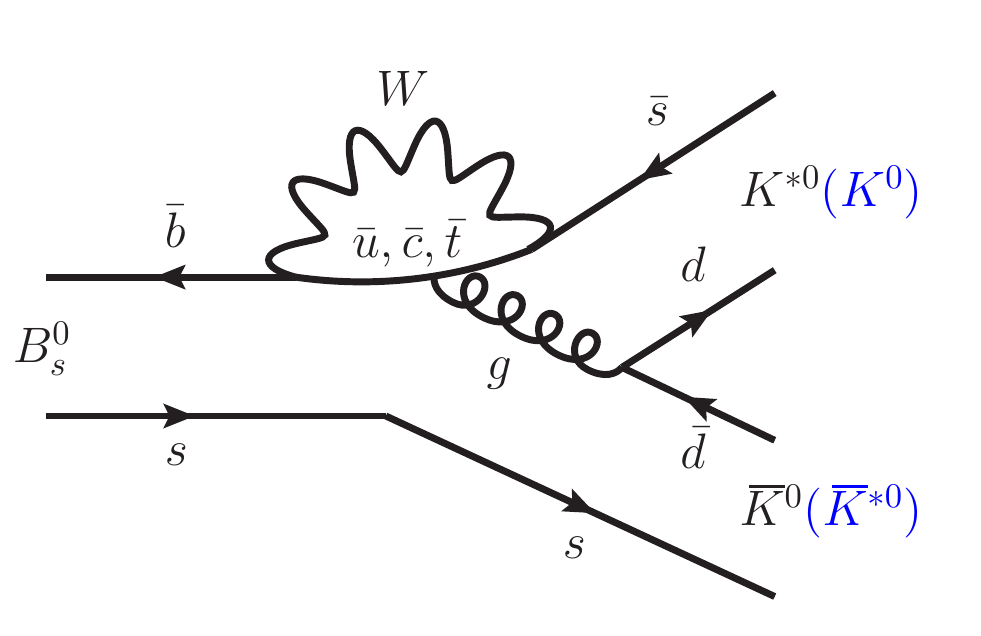}
  \caption{\small
    Feynman diagrams for (top left) external tree, (top right) internal penguin and 
    (bottom left) electroweak penguin contributions for $\Bs \to \Kp\Kstarm~(\Kstarp\Km)$ decays; and 
    (bottom right) internal penguin amplitude for the $\Bs \to \Kstarz\Kzb~(\Kz\Kstarzb)$ decay mode. 
    The electroweak penguin diagram for the $\Bs \to \Kstarz\Kzb~(\Kz\Kstarzb)$ channel is not shown; neither are diagrams corresponding to annihilation amplitudes. 
    In each case, the first set of final-state particles (black) leads to the $\KS\Kp\pim$ final state, while the second set (blue) leads to $\KS\Km\pip$.
  }
  \label{fig:feynman}
\end{figure}

The $\Bs\to\KS\Kpm\pimp$ channels have been observed~\cite{LHCb-PAPER-2013-042,LHCb-PAPER-2017-010}, and quasi-two-body measurements of the resonant contributions from $\Bs\to\Kstarpm\Kmp$~\cite{LHCb-PAPER-2014-043} and $\Bs\to\KstarzorKstarzb\KS$~\cite{LHCb-PAPER-2015-018} decays have also been performed.
These decays provide interesting potential for time-dependent \CP-violation measurements~\cite{Gronau:2006qn}, once sufficiently large samples become available.
The $\KS\Km\pip$ and $\KS\Kp\pim$ final states are not flavour-specific and as such both \Bs and \Bsb decays can contribute to each, with the corresponding amplitudes expected to be comparable in magnitude.
Large interference effects and potentially large \CP-violation effects are possible, making an amplitude analysis of these channels of particular interest.
Example decay diagrams for contributions through the $\Bs \to \Kp\Kstarm~(\Kstarp\Km)$ and $\Bs \to \Kstarz\Kzb~(\Kz\Kstarzb)$ resonant processes are shown in Fig.~\ref{fig:feynman}.
The subsequent transitions $\Kstarm \to \Kzb\pim$, $\Kstarz \to \Kp\pim$ and $\Kzb \to \KS$ (and their conjugates) lead to the $\KS\Kp\pim$ ($\KS\Km\pip$) final state for the former (latter) processes.\footnote{
The inclusion of charge conjugate processes is implied throughout the paper, except where explicitly stated otherwise.}

In this article, the first Dalitz plot analysis of $\Bs\to\KS\Kpm\pimp$ decays is described.  
The analysis is based on a sample corresponding to $3.0\invfb$ of $pp$ collision data recorded with the LHCb detector during 2011 and 2012.
Due to the limited signal yield, and the modest effective tagging efficiency that can be achieved at hadron collider experiments, the analysis is performed without considering decay-time dependence and without separating the $\Bs$ or $\Bsb$ initial states (\ie\ the analysis is \emph{untagged}).
As such, the current analysis has limited sensitivity to \CP-violation effects but provides an important basis for future studies. 

A novel feature of this analysis is that there are two independent final states ($\KS\Kp\pim$ and $\KS\Km\pip$) that are treated separately but simultaneously.
Denoting one final state by $f$ and the other by $\bar{f}$, the former (latter) receives contributions from the amplitudes ${\cal{A}}_f$ and $\bar{\cal{A}}_f$ (${\cal{A}}_{\bar{f}}$ and $\bar{\cal{A}}_{\bar{f}}$), where ${\cal{A}}$ and $\bar{\cal{A}}$ are used to denote amplitudes for \Bs\ and \Bsb\ decays, respectively.
Therefore, the untagged decay-time-integrated density of events in the Dalitz plot corresponding to $f$ depends on $|{\cal A}_{f}|^2$ and $|\bar{{\cal A}}_{f}|^2$, while that for $\bar{f}$ depends on $|{\cal A}_{\bar{f}}|^2$ and $|\bar{{\cal A}}_{\bar{f}}|^2$.
The untagged decay-time-integrated rate also depends on an interference term that is responsible for the difference between the decay probability at $t=0$ and the decay-time-integrated branching fraction~\cite{DeBruyn:2012wj,LHCb-PAPER-2013-069,Dettori:2018bwt}.
This must be considered when results are interpreted theoretically, but is not relevant for the discussion here.
In the absence of \CP\ violation in decay ${\cal{A}}_f = {\bar{\cal{A}}}_{\bar{f}}$ and ${\bar{\cal{A}}}_f = {\cal{A}}_{\bar{f}}$, but there is no simple relation between ${\cal{A}}_f$ and ${\bar{\cal{A}}}_f$.
Indeed, theoretical predictions indicate that the values of these amplitudes could be quite different~\cite{Cheng:2014uga,Li:2014fla,Li:2018qrm}.
Thus, the situation differs from that usually considered in Dalitz plot analyses, where the density is given by just the magnitude of a single amplitude squared. 

A precedent for handling this situation is taken from amplitude analyses of flavour-specific \B-meson decays that do not account for \CP-violation effects.  
In such analyses the distributions for $B$ and $\Bbar$ decays are summed, assuming them to be identical, so that they can be fitted with a single amplitude.
However, in the presence of \CP-violation effects, the distribution is actually given by the incoherent sum of two contributions, as is the case here.  
Consequently, the fitted parameters of the amplitude model will differ from their true values by an amount that depends on the size of the \CP-violation effects.
Similarly, by fitting each of the two $\Bs\to\KS\Kpm\pimp$ Dalitz plots with a single amplitude, the results will give values that differ from the true properties of the decays by amounts that must be estimated.
Detailed studies with simulated pseudoexperiments demonstrate that the fit fractions (defined in Sec.~\ref{sec:formalism}) obtained by this approach are biased by relatively small amounts that can be accounted for with systematic uncertainties, but that measurements of other quantities may not be reliable.
Therefore, the results of the analysis are presented in terms of fit fractions only.

The remainder of the paper is organised as follows.
In Sec.~\ref{sec:Detector}, a brief description of the LHCb detector, online selection algorithms and simulation software is given.
The selection of $\Bs\to\KS\Kpm\pimp$ candidates, and the method to estimate the signal and background yields are described in Sec.~\ref{sec:selection} and Sec.~\ref{sec:dataset}, respectively.
The analysis described in these sections follows closely the methods used for the branching fraction measurement presented in Ref.~\cite{LHCb-PAPER-2017-010}.
As such, all four final states (\KsPiPi, \KsKpPim, \KsKmPip, and \KsKK, collectively referred to as $\Kshhp$ where $h$ represents either a kaon or a pion) are considered up to Sec.~\ref{sec:dataset},
where the inclusion of the \KsPiPi\ and \KsKK\ modes aids control of backgrounds due to misidentified final-state particles.
Only the $\KsKpPim$ and $\KsKmPip$ channels are discussed subsequently in the paper.
The Dalitz plot analysis formalism is presented in Sec.~\ref{sec:formalism} and inputs to the fit such as the signal efficiency and background distributions are described in Sec.~\ref{sec:dalitz}.
Sources of systematic uncertainty are discussed in Sec.~\ref{sec:systematics}, before the results are presented in Sec.~\ref{sec:results}.
A summary concludes the paper in Sec.~\ref{sec:summary}.

\section{Detector, trigger and simulation}
\label{sec:Detector}

The \lhcb detector~\cite{Alves:2008zz,LHCb-DP-2014-002} is a single-arm
forward spectrometer covering the \mbox{pseudorapidity} range $2<\eta <5$,
designed for the study of particles containing \bquark or \cquark quarks.
The detector includes a high-precision tracking system consisting of a
silicon-strip vertex detector (\velo) surrounding the $pp$ interaction
region, a large-area silicon-strip detector located upstream of a dipole
magnet with a bending power of about $4{\mathrm{\,Tm}}$, and three stations
of silicon-strip detectors and straw drift tubes placed downstream of the
magnet.
The tracking system provides a measurement of the momentum, \ptot, of charged particles with relative uncertainty that varies from 0.5\% at low momentum to 1.0\% at 200\gevc.
The minimum distance of a track to a primary vertex (PV), the impact parameter, is measured with a resolution of $(15+29/\pt)\mum$,
where \pt is the component of the momentum transverse to the beam, in\,\gevc.
Different types of charged hadrons are distinguished using information
from two ring-imaging Cherenkov detectors.
Photons, electrons and hadrons are identified by a system
consisting of scintillating-pad and preshower detectors, and electromagnetic and hadronic calorimeters.
Muons are identified by a system composed of alternating layers of iron and
multiwire proportional chambers.

The online event selection is performed by a
trigger~\cite{LHCb-DP-2012-004}, 
which consists of a hardware stage, based on information from the calorimeter and muon
systems, followed by a software stage, in which all charged particles
with $\pt>500\,(300)\mevc$ are reconstructed for data collected in 2011\,(2012).
At the hardware trigger stage, events are required to contain a muon with high
\pt or a hadron, photon or electron with high transverse energy in the
calorimeters.
The software trigger requires a two-, three- or four-track secondary vertex
with significant displacement from all primary $pp$ interaction vertices.
At least one charged particle must have $\pt >
1.7\,(1.6)\gevc$ in the 2011\,(2012) data and be inconsistent with
originating from a PV.
A multivariate algorithm~\cite{BBDT} is used for the identification of
secondary vertices consistent with the decay of a \bquark hadron.
It is required that the software trigger decision must have been caused
entirely by tracks from the decay of the signal \B candidate.

Simulated data samples are used to investigate backgrounds from other
\bquark-hadron decays and also to study the detection and reconstruction
efficiency of the signal.
In the simulation, $pp$ collisions are generated using
\pythia~\cite{Sjostrand:2007gs,*Sjostrand:2006za} with a specific \lhcb
configuration~\cite{LHCb-PROC-2010-056}.
Decays of hadronic particles are described by \evtgen~\cite{Lange:2001uf},
in which final-state radiation is generated using
\photos~\cite{Golonka:2005pn}.
The interaction of the generated particles with the detector, and its
response, are implemented using the \geant toolkit~\cite{Allison:2006ve,
*Agostinelli:2002hh} as described in Ref.~\cite{LHCb-PROC-2011-006}.

\section{Event selection}
\label{sec:selection}

The selection requirements follow closely those used for the determination
of the branching fractions of the \BdstoKshhp decays, reported in
Ref.~\cite{LHCb-PAPER-2017-010}.
A brief summary of the requirements follows, with emphasis placed on
where they differ from those used in the branching-fraction analysis.

Decays of \decay{\KS}{\pip\pim} are reconstructed in two categories:
the first involving \KS mesons that decay early enough for the
resulting pions to be reconstructed in the \velo; and the
second containing \KS mesons that decay later, such that track
segments from the pions cannot be formed in the \velo.
These categories are referred to as \emph{\LL} and \emph{\DD}, respectively.
While the \LL\ category has better mass, momentum and vertex resolution,
there are approximately twice as many \KS candidates reconstructed in the
\DD\ category.
In the following, \B candidates reconstructed from either a \LL\ or \DD\
\KS candidate, in addition to two oppositely charged tracks, are also
referred to with these category names.
In order to account for changes in the trigger efficiency for each of the
\KS reconstruction categories during the data taking, the data sample is
subdivided into 2011, 2012a, and 2012b data-taking periods.
The 2012b sample is the largest, corresponding to 1.4\invfb, and also has
the highest trigger efficiency.

To suppress backgrounds, in particular combinatorial background formed from
random combinations of unrelated tracks, the events satisfying the trigger
requirements are filtered by a loose preselection, followed by a
multivariate selection optimised separately for each data sample.
All requirements are made with care to minimise correlation of the
signal efficiency with position in the Dalitz plot, resulting in better
control of the corresponding systematic uncertainties.
Consequently, the selection relies very little on the kinematics of the final-state particles and instead exploits heavily the topological features that arise from the detached vertex of the \B candidate. 
These include: the impact parameters of the \B candidate and its decay products, the quality of the decay vertices of the \B and \KS candidates, as well as the separation of these vertices from each other and from the primary vertex, and their isolation from other tracks in the event.

The preselection of \KS and \B candidates and the training of the
multivariate classifiers, based on a boosted decision tree~(BDT)
algorithm~\cite{Breiman,AdaBoost}, is identical to that reported in
Ref.~\cite{LHCb-PAPER-2017-010}.
The selection requirement placed on the output of each of the BDTs is optimised using the figure of merit
\begin{equation}
  {\cal Q} \equiv \frac{N^2_{\rm sig}}{\left(N_{\rm sig}+N_{\rm bkg}\right)^\frac{3}{2}} \,,
  \label{eq:StimesPFom}
\end{equation}
where $N_{\rm sig}$ ($N_{\rm bkg}$) represents the expected signal
(combinatorial background) yield in the combined $\KS\Kpm\pimp$ sample, for a given selection, in the signal region defined in Sec.~\ref{sec:dataset}.
This figure of merit, which is different from that in Ref.~\cite{LHCb-PAPER-2017-010}, is found to be suitable for Dalitz plot analyses in a dedicated study.
Pseudoexperiments are generated using a model containing a set of resonances that might contribute to the \BstoKsKPi Dalitz plot, and signal and background yields corresponding to various possible selection requirements on the BDT output.
The statistical uncertainty on each of the magnitudes and phases of the
resonances in the model, as well as the systematic uncertainty
corresponding to the knowledge of the Dalitz plot distribution of the
backgrounds, are determined for each selection requirement.
The responses of several figures of merit are compared with the results of this study, and that given in Eq.~\eqref{eq:StimesPFom} is found to show the closest correspondence to minimising the uncertainties on the amplitude parameters.
It may be noted that ${\cal Q}$ is equal to the product of two other figures of merit considered in the literature: $N_{\rm sig}/\sqrt{N_{\rm sig}+N_{\rm bkg}}$ (sometimes referred to as \emph{significance}) and $N_{\rm sig}/\left(N_{\rm sig}+N_{\rm bkg}\right)$ (\emph{purity}).

Particle identification (PID) information is used to assign
each candidate exclusively to one of the four possible final states:
\KsPiPi, \KsKpPim, \KsKmPip, and \KsKK.
The PID requirements are optimised to reduce the cross-feed between the
different signal decay modes using the same figure of merit ${\cal Q}$ introduced for the BDT optimisation. 
Additional PID requirements are applied in order to reduce backgrounds from
decays such as \LbtoKsPip, where the proton is misidentified as a kaon.

Fully reconstructed \B-meson decays into two-body $\DorDsm\hadp$ or
$(\cquark\cquarkbar)\KS$ combinations, where $(\cquark\cquarkbar)$
indicates a charmonium resonance, may result in a \Kshhp final state that
satisfies the selection criteria and has the same \B-candidate
invariant-mass distribution as the signal candidates.
The decays of \Lbbar\ baryons to $\Lcbar\hadp$ with \decay{\Lcbar}{\antiproton\KS} also
peak under the signal when the antiproton is misidentified.
A series of invariant-mass vetoes, identical to those used in
Ref.~\cite{LHCb-PAPER-2017-010}, are employed to remove these backgrounds.

Less than 1\% of selected events contain more than one \B candidate.
The candidate that is retained in such events is chosen in a random but reproducible manner.

\section{Determination of signal and background yields}
\label{sec:dataset}

The signal and background yields are determined by means of a simultaneous
unbinned extended maximum-likelihood fit to the 24 \B-candidate invariant-mass distributions that result from considering separately the four final
states, three data-taking periods and two \KS reconstruction categories.
Three components contribute to each invariant-mass distribution:
signal decays, backgrounds resulting from cross-feeds, and random
combinations of unrelated tracks.
The contribution from a fourth category of background, partially
reconstructed decays, is reduced to a negligible level by performing the
fit in the invariant-mass window $5200 < m(\Kshhp) < 5800\mevcc$.
The modelling of each of the three fit components follows that used in
Ref.~\cite{LHCb-PAPER-2017-010}.
A brief summary of the models used is given here.

The \B-candidate mass distributions for signal decays with correctly identified final-state particles are modelled with the sum of two Crystal Ball (CB)
functions~\cite{Skwarnicki:1986xj} that share common values for the mean
and width of the Gaussian part of the function but have independent power-law tails on opposite sides of the Gaussian peak.
Cross-feed contributions from misidentified $\BdstoKshhp$ decays are also modelled with the sum of two CB functions.
Only processes with a single misidentified track are included, since other potential misidentified decays are found to have negligibly small contributions.
The yield of each misidentified decay is constrained, with respect to the
yield of the corresponding correctly identified decay, using the ratio of
the selection efficiencies and the corresponding uncertainty.
The combinatorial background is modelled by an exponential function.

\begin{figure}[!tb]
\begin{center}
\includegraphics*[width=0.49\textwidth]{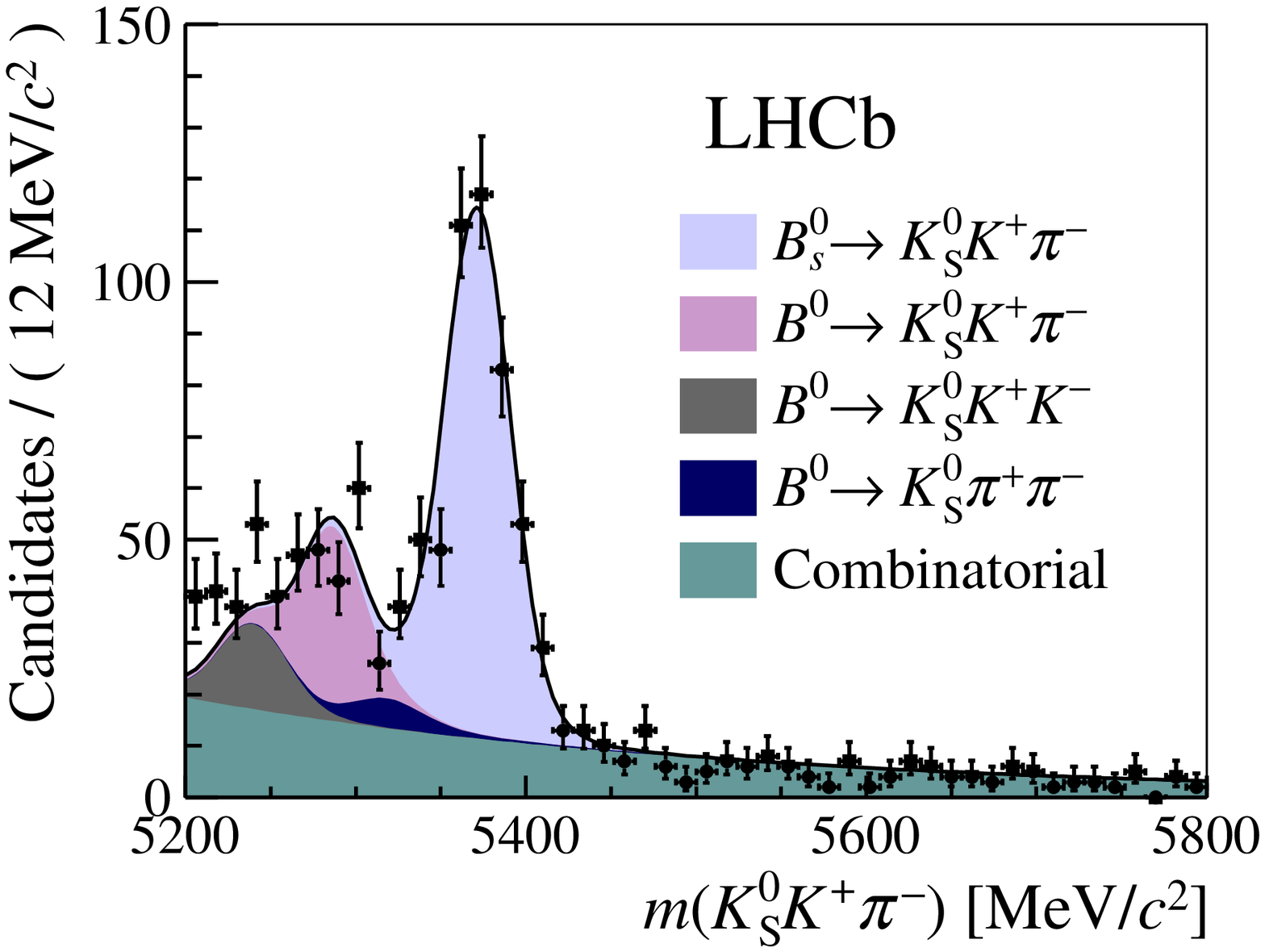}
\includegraphics*[width=0.49\textwidth]{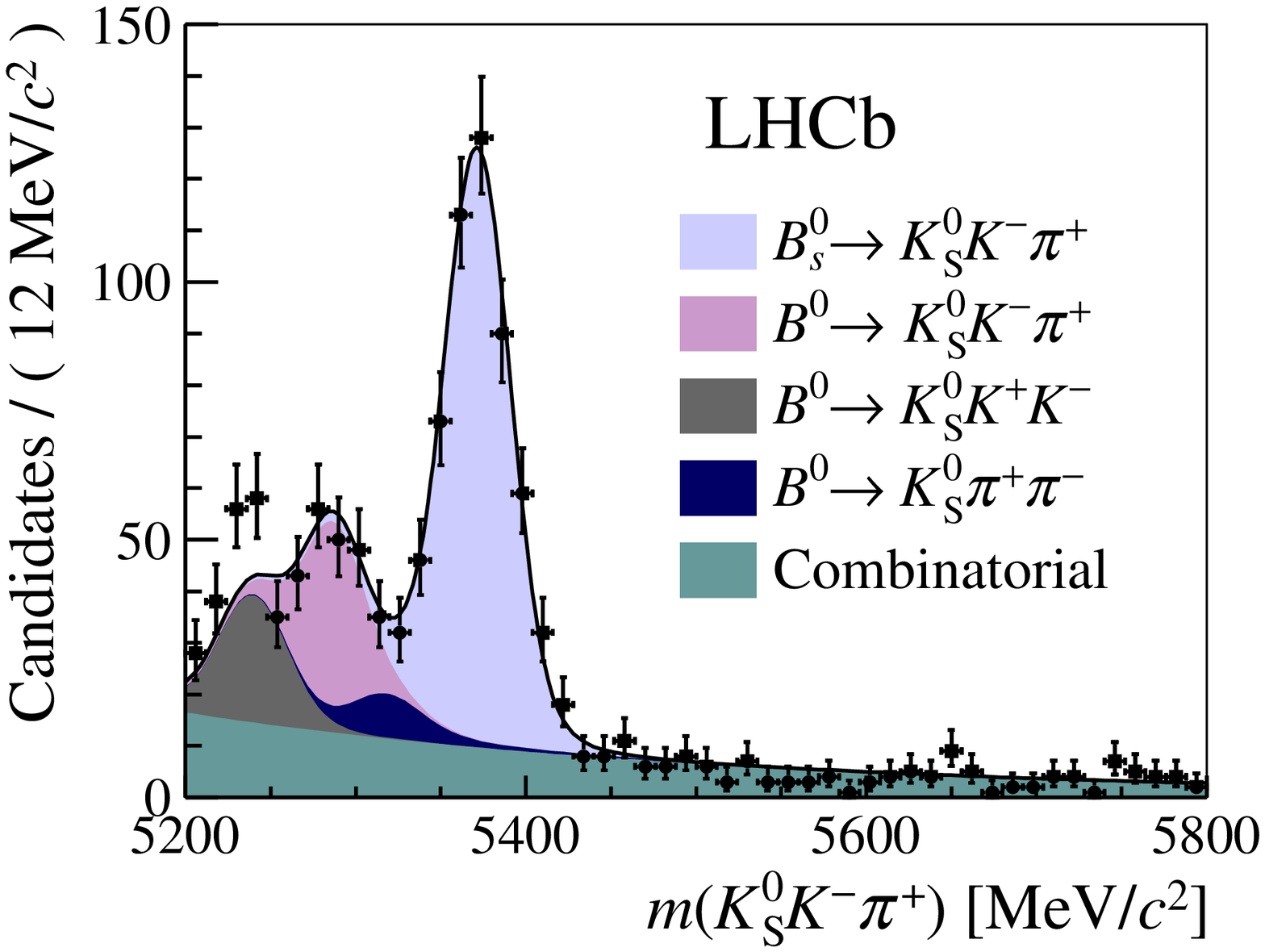}
\end{center}
\caption{\small
  Invariant-mass distribution of candidates in data for the (left)~\KsKpPim\ and (right)~\KsKmPip\ final states. 
  Components are detailed in the legend.
}
\label{fig : mass-fit}
\end{figure}

The fit results for the \KsKpPim and \KsKmPip final states,
combining all data-taking periods and \KS reconstruction categories,
are shown in \fig{mass-fit}, 
where comparison of the data with the result of the fit gives $\chisq$ values of 49.6 and 35.3 for the 50 mass bins in each of the \KsKpPim and \KsKmPip final states.\footnote{
  Since \fig{mass-fit} contains projections of the simultaneous fit to 24 invariant-mass distributions, the numbers of degrees of freedom associated to these $\chisq$ values cannot be trivially calculated.
}
\tab{mass-fit} details the fitted yields of all subsamples of the \KsKpPim and \KsKmPip final states, both in the
invariant-mass region used for the mass fit and in the reduced region to be
used in the amplitude analysis, defined as $\mu\pm2.5\sigma$ where $\mu$
($\sigma$) is the fitted peak position (width) of the \Bs signal component in
that category.
The yields are given for each of the two final states split by data-taking periods and \KS reconstruction categories.
Within the reduced region used in the amplitude analysis, a total of $529$ and $573$ candidates are selected for the $\KS K^+\pi^-$ and $\KS K^-\pi^+$ final states, respectively.

\begin{table}[t]
\caption{\small
  Yields obtained from the simultaneous fit to the invariant-mass
  distribution of \mbox{$\Bs\to\KS\Kpm\pimp$} candidates in data for each fit category: signal, combinatorial background and cross-feed from misidentified $\BdstoKshhp$ decays.
  The uncertainties given on the yields in the full range are statistical only.
  Yields in the signal region $\pm2.5\sigma$ around the \Bs\ peak are also given; the determination of uncertainties on these values is described in Sec.~\ref{sec:systematics}.
}
\label{tab : mass-fit}
\resizebox{\textwidth}{!}{
\begin{tabular}{ c | c | c | c  c | c  c | c c }
\hline
  Final			& \KS		& Sample	& \multicolumn{2}{c}{\Bs\ signal}		& \multicolumn{2}{c}{Combinatorial}		& \multicolumn{2}{c}{Cross-feed}	\\
state			& category	& & Full range			& $2.5\,\sigma$	& Full range			& $2.5\,\sigma$	& Full range		& $2.5\,\sigma$	\\
\hline                                                                                                                                                                                  
\mrsix{\KsKpPim}	& \mrthree{\DD}	& 2011	& $\pho73.6 \pm 10.6$		& $\pho72.1$	& $108.3 \pm 15.1$		& $22.1$	& $\pho8.9 \pm 2.8$	        & $\pho1.7$		\\
			&		& 2012a	& $\pho48.2 \pm \pho8.6$	& $\pho45.7$	& $\pho70.1 \pm 12.1$		& $14.3$	& $\pho7.3 \pm 3.8$	        & $\pho1.1$		\\
			&		& 2012b	& $135.3 \pm 13.6$		& $130.0$	& $\pho87.4 \pm 13.8$		& $17.9$	& $17.0 \pm 5.6$	& $\pho3.1$		\\
\cdashline{2-9}                                                                                                                                                                         
			& \mrthree{\LL}	& 2011	& $\pho76.2 \pm \pho9.8$	& $\pho74.6$	& $\pho44.1 \pm \pho9.8$	& $\pho8.4$	& $\pho8.2 \pm 1.7$	        & $\pho1.8$		\\
			&		& 2012a	& $\pho38.5 \pm \pho7.7$	& $\pho36.8$	& $\pho58.8 \pm 11.2$		& $11.2$	& $\pho7.8 \pm 1.8$	        & $\pho0.9$		\\
			&		& 2012b	& $\pho73.5 \pm 10.6$		& $\pho71.9$	& $\pho71.7 \pm 13.1$		& $13.6$	& $15.9 \pm 2.5$	& $\pho1.7$		\\
\cdashline{2-9}                                                                                                                                                                         
			& total	&	&				& 431.1		&				& 87.5		&			& 10.3		\\
\hline                                                                                                                                                                                  
\mrsix{\KsKmPip}	& \mrthree{\DD}	& 2011	& $\pho72.8 \pm 10.3$		& $\pho71.4$	& $\pho78.9 \pm 12.7$		& $16.1$	& $\pho8.2 \pm 2.4$    	& $\pho1.3$		\\
			&		& 2012a	& $\pho68.8 \pm \pho9.6$	& $\pho65.2$	& $\pho46.2 \pm \pho9.9$	& $\pho9.5$	& $\pho7.0 \pm 3.4$	        & $\pho1.2$		\\
			&		& 2012b	& $165.1 \pm 15.2$		& $158.6$	& $104.1 \pm 15.0$		& $21.3$	& $17.3 \pm 5.8$	& $\pho2.9$		\\
\cdashline{2-9}                                                                                                                                                                         
			& \mrthree{\LL}	& 2011	& $\pho77.3 \pm \pho9.8$	& $\pho75.7$	& $\pho39.0 \pm 10.2$		& $\pho7.4$	& $\pho9.6 \pm 1.7$	        & $\pho1.4$		\\
			&		& 2012a	& $\pho40.3 \pm \pho8.1$	& $\pho38.5$	& $\pho58.9 \pm 11.9$		& $11.2$	& $\pho8.6 \pm 1.8$	        & $\pho0.7$		\\
			&		& 2012b	& $\pho81.7 \pm 10.4$		& $\pho80.0$	& $\pho50.1 \pm 12.3$		& $\pho9.5$	& $15.0 \pm 2.5$	& $\pho1.4$		\\
\cdashline{2-9}                                                                                                                                                                         
			& total	&	&				& 489.4		&				& 75.0		&			& $\pho8.9$		\\
\hline
\end{tabular}
}
\end{table}

%
\section{Dalitz plot analysis formalism}
\label{sec:formalism}

The Dalitz plot~\cite{Dalitz:1953cp} describes the phase-space of a three-body decay in terms of two of the three possible two-body invariant-mass squared combinations.
In $\Bs \to \KS\Kpm\pimp$ decays the most significant resonant structures are expected to be from excited kaon states decaying to $\KS\pimp$ or $\Kpm\pimp$ and therefore these are used to define the Dalitz plot axes.
The values of $m(\KS\pimp)$ and $m(\Kpm\pimp)$ are calculated following a kinematic fit~\cite{Hulsbergen:2005pu} in which the \Bs candidate mass is fixed to the known value of $m_{\Bs}$~\cite{PDG2018}, which improves resolution and ensures that all decays remain within the Dalitz plot boundary.
These values and $m_{\Bs}$ are used to calculate all other kinematic quantities that are used in the Dalitz plot fit.

The Dalitz plot analysis involves developing a model that describes the variation of the complex decay amplitudes over the full phase-space of a three-body decay.
The observed distribution of decays is related to the square of the magnitude of the amplitude, modified to account for detection efficiency and background contributions.  
As described in Sec.~\ref{sec:Introduction}, this is only an approximation for $\Bs \to \KS\Kpm\pimp$ decays, where the physical distribution in each final state depends on the incoherent sum of two contributions.  
A single amplitude is nonetheless used to model the data, since it is not possible to separate the two contributing amplitudes without initial-state flavour tagging; a systematic uncertainty is assigned to account for possible biases induced by this approximation.
The Dalitz plot fit is performed using the {\sc Laura++}~\cite{Laura++} package, with the different final states, \KS reconstruction categories and data-taking periods handled using the {\it J}{\sc fit} method~\cite{Ben-Haim:2014afa}.

The isobar model~\cite{Fleming:1964zz,Morgan:1968zza,Herndon:1973yn} is used to describe the complex decay amplitude. 
The total amplitude is given by the coherent sum of $N$ intermediate processes,
\begin{equation}\label{eqn:amp}
  {\cal A}\!\left[m^2(\KS\pimp), m^2(\Kpm\pimp)\right] = \sum_{j=1}^{N} c_j F_j\!\left[m^2(\KS\pimp), m^2(\Kpm\pimp)\right] \,,
\end{equation}
where $c_j$ are complex coefficients describing the relative contribution of each intermediate amplitude. 
The resonant dynamics are contained in the $F_j\!\left[m^2(\KS\pimp),m^2(\Kpm\pimp)\right]$ terms, which are normalised such that the integral of the squared magnitude over the Dalitz plot is unity for each term.
For a $\KS\pimp$ resonance $F_j\!\left[m^2(\KS\pimp),m^2(\Kpm\pimp)\right]$ is given by
\begin{equation}
  \label{eq:ResDynEqn}
  F\!\left[m^2(\KS\pimp), m^2(\Kpm\pimp)\right] = 
  R\!\left[m(\KS\pimp)\right] \times X(|\vec{p}\,|\,r_{\rm BW}) \times X(|\vec{q}\,|\,r_{\rm BW}) 
  \times T(\vec{p},\vec{q}\,) \, ,
\end{equation}
where $\vec{p}$ is the momentum of the \emph{companion} particle\footnote{
  The companion particle is that not forming the resonance, \ie\ the $\Kpm$ in this example.} and $\vec{q}$ is the momentum of one of the resonance decay products, both evaluated in the $\KS\pimp$ rest frame.
The $R$ functions are the mass lineshapes, typically described by the relativistic Breit--Wigner function with alternative shapes used in some specific cases.
The $X$ and $T$ terms describe barrier factors and angular distributions, respectively, and depend on the orbital angular momentum between the resonance and the companion particle, $L$.
The barrier factors $X$ are evaluated in terms of the Blatt--Weisskopf radius parameter $r_{\rm BW}$ for which a default value of $4.0 \gev^{-1}\hbar c$ is used.
The angular distributions are given in the Zemach tensor formalism~\cite{Zemach:1963bc,Zemach:1968zz}, and are proportional to the Legendre polynomials, $P_L(x)$, where $x$ is the cosine of the angle between $\vec{p}$ and $\vec{q}$ (referred to as the helicity angle). 
Detailed expressions for the functions $R$, $X$ and $T$ can be found in Ref.~\cite{Laura++}.

The complex coefficients $c_j$, defined in Eq.~\eqref{eqn:amp}, are determined from the fit to data.
These are used to obtain fit fractions for each component $j$, which provide a robust and convention-independent way to report the results of the analysis.
The fit fractions are defined as the integral over one Dalitz plot ($\KS\Kp\pim$ or $\KS\Km\pip$) of the amplitude for each intermediate component squared, divided by that of the coherent matrix element squared for all intermediate contributions, 
\begin{equation}
{\cal F}_j =
\frac
{\int\!\!\int_{\rm DP}\left|c_j F_j\right|^2~{\rm d}m^2(\KS\pimp)\,{\rm d}m^2(\Kpm\pimp)}
{\int\!\!\int_{\rm DP}\left|{\cal A}\right|^2~{\rm d}m^2(\KS\pimp)\,{\rm d}m^2(\Kpm\pimp)} \, ,
\label{eq:fitfraction}
\end{equation}
where the dependence of $F_j$ and ${\cal A}$ on Dalitz plot position has been omitted for brevity.
The fit fractions need not sum to unity due to possible net constructive or destructive interference.

For this analysis, it is useful to define also flavour-averaged fit fractions $\widehat{{\cal F}_{j}}$, where the numerator and denominator of Eq.~\eqref{eq:fitfraction} are replaced by sums of the same quantities over both final states, and it is understood that a resonance corresponding to $j$ in one Dalitz plot will be replaced by its conjugate in the other (\eg\ \KstarIm\ in the $\KS\Kp\pim$ final state and \KstarIp\ for $\KS\Km\pip$).
These can be converted into products of branching fractions for the \Bs\ and \Kstar\ decays by multiplying by the known $\BsToKzBarOptKpi$ branching fraction,
\begin{equation}
  \label{eq:Chap-FFhat}
  \Br{\Bs \to \Kstar\kaon; \Kstar\to\kaon\pion} = \widehat{{\cal F}_{j}} \times \Br{\BsToKzBarOptKpi} \,, 
\end{equation}  
where $\Br{\Bs \to \Kstar\kaon}$ is the sum of the branching fractions for the two conjugate final states.

\section{Dalitz plot fit}
\label{sec:dalitz}

The parameters of the signal model are determined from a simultaneous unbinned maximum-likelihood fit to the distribution of data across the $\KS\Kp\pim$ and $\KS\Km\pip$ Dalitz plots.  
The physical signal model is modified to account for variation of the efficiency across the phase-space, and background contributions are included.
The yields of signal and background components in the signal region are taken from Table~\ref{tab : mass-fit}.
Separate efficiency functions and background models for each final state, \KS reconstruction category and data-taking period are also used.

Since the resonance masses are much smaller than the \Bs\ mass, the selected candidates tend to populate regions close to the kinematic boundaries of the Dalitz plot.
Therefore, it is convenient to describe the signal efficiency variation and background event density using the transformed coordinates referred to as square Dalitz plot (SDP) variables.
The SDP variables are defined by
\begin{equation}
\label{eq:sqdp-vars}
m^{\prime} \equiv \frac{1}{\pi}
\arccos\left(2\frac{m(\Kpm\pimp) - m^{\rm min}_{\Kpm\pimp}}{m^{\rm max}_{\Kpm\pimp} - m^{\rm min}_{\Kpm\pimp}} - 1 \right)\,, 
\qquad
\theta^{\prime} \equiv \frac{1}{\pi}\theta(\Kpm\pimp)\,,
\end{equation}
where $m(\Kpm\pimp)$ is the invariant mass of the charged kaon and pion,
$m^{\rm max}_{\Kpm\pimp} = m_{\Bs} - m_{\KS}$ and
$m^{\rm min}_{\Kpm\pimp} = m_{\Kpm} + m_{\pimp}$
are the kinematic limits of $m_{\Kpm\pimp}$,
and $\theta(\Kpm\pimp)$ is the helicity angle between the \pimp\ and the \KS\ in the $\Kpm\pimp$ rest frame.

\subsection{Signal efficiency variation}

Variation across the phase space of the probability to reconstruct a signal decay is accounted for in the fit by multiplying the amplitude squared by an efficiency function~\cite{Laura++}.
The signal efficiency is determined including effects due to the LHCb detector geometry, and due to reconstruction and selection requirements.
The effects of PID requirements are considered separately to the rest of the selection efficiency to facilitate their determination using data control samples. 

The geometric efficiency is determined from generator-level simulation.
This contribution is the same for the 2012a and 2012b samples, and for the \LL and \DD categories, as it is purely related to the kinematics of \Bs\ mesons that are produced in $pp$ collisions at the LHC.
The effect is evaluated separately for the 2011 and 2012 data due to the different beam energies.

The product of the reconstruction and selection (excluding PID) efficiencies is determined from simulated samples, which account for the response of the detector, generated with a flat distribution across the square Dalitz plot.
Small corrections are applied to take into account known differences between data and simulation in the track-finding efficiency~\cite{DeCian:1402577} and hardware-trigger response~\cite{MartinSanchez:1407893}.

The efficiency of the PID requirements is determined from large control samples of $\Dstarp \to \Dz \pip$, $\Dz \to \Km\pip$ decays.
Differences in kinematics and detector occupancy between the control samples and the signal data are accounted for~\cite{LHCb-DP-2012-003,LHCb-PUB-2016-021}.

The combined efficiency maps are obtained as products of SDP histograms describing each of the three contributions described above.
These are subsequently smoothed using two-dimensional bicubic splines.
The variation of the efficiency across the SDP is similar for each subsample of the data; the absolute scale differs between \LL and \DD categories due to acceptance and between data-taking periods due to changes in the trigger.
The efficiency varies by about a factor of five 
between the smallest and largest values, mainly caused by the difficulty to reconstruct decays in a region of phase-space where the $\Kpm$ and $\pimp$ tracks have low momentum and the $\KS$ is highly energetic.

\subsection{Background modelling}

As can be seen in Fig.~\ref{fig : mass-fit} and Table~\ref{tab : mass-fit}, the signal region contains contributions from combinatorial background and cross-feed from misidentified $\Bz\to\KS\pip\pim$ decays.
The Dalitz plot distribution of the combinatorial background is modelled using data from a sideband at high $m(\KS\Kpm\pimp)$. 
In order to increase the size of the sample used for this modelling, a looser BDT requirement is imposed than that used for the signal selection.
It is verified that this does not change the Dalitz plot distribution of the background significantly (the BDT is explicitly constructed to minimise correlation of its output variable with position in the Dalitz plot).
The combinatorial background is found to vary smoothly over the Dalitz plot.

Cross-feed from misidentified $\Bz\to\KS\pip\pim$ decays is modelled using a simulation of this decay, weighted in order to reproduce its measured Dalitz plot distribution~\cite{Aubert:2009me}.
The effect of the detector response is simulated, with the effect of the PID requirements accounted for by weights determined from data control samples, as is done for the evaluation of the signal efficiency.  
The most prominent structures in the Dalitz plot model for this background are due to the \KstarIpm\ resonances.

\subsection{Amplitude model for \boldmath{$\Bs \to \KS\Kpm\pimp$} decays} 

\begin{figure}[!tb]
  \begin{center}
    \includegraphics*[width=0.49\textwidth]{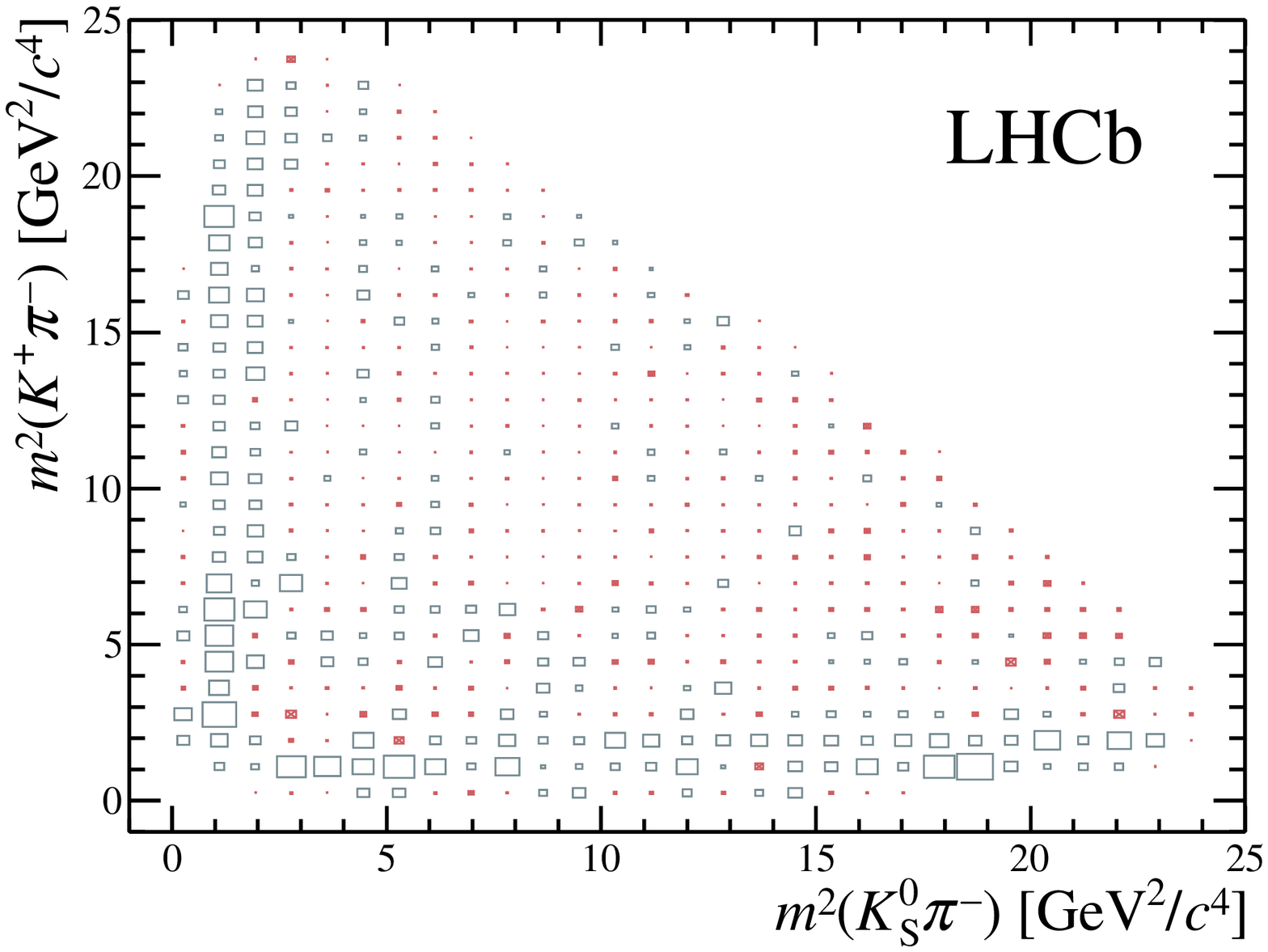}
    \includegraphics*[width=0.49\textwidth]{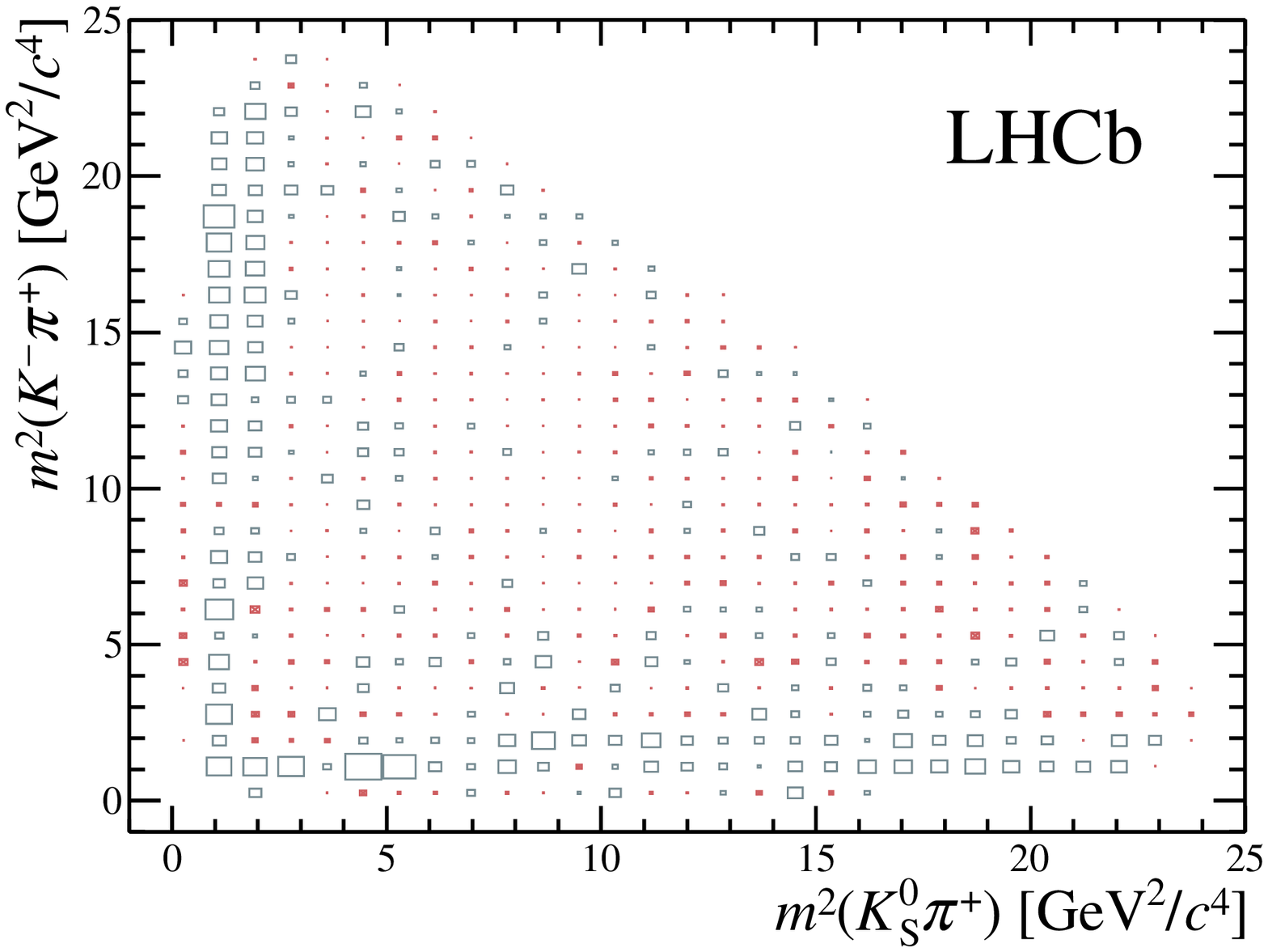}
  \end{center}
\caption{\small
  Background-subtracted and efficiency-corrected Dalitz plot distributions for (left)~\KsKpPim and (right)~\KsKmPip final states. 
  Boxes with a cross indicate negative values.
  }
  \label{fig:dp-distribution}
\end{figure}

The Dalitz plot distributions of the selected $\Bs\to\KS\Kpm\pimp$ candidates, after background subtraction and efficiency correction, are shown in Fig.~\ref{fig:dp-distribution} for all data subsamples combined.
There are clear excesses at low values of both $m^2(\KS\pimp)$ and $m^2(\Kpm\pimp)$, corresponding to excited kaon resonances.
There is no strong excess at low values of $m^2(\KS\Kpm)$, which would appear as diagonal bands towards the upper right side of the kinematically allowed regions of the Dalitz plots.
The two Dalitz plot distributions appear to be consistent with each other, and hence with \CP\ conservation.

The baseline signal model is developed by assessing the impact of including or removing resonant or nonresonant contributions in the model.
The kaon resonances listed in Ref.~\cite{PDG2018} are considered.
Charged and neutral isospin partners are treated separately, as it is possible that one contributes significantly while the other does not.
If a resonance is included in the model for one final state, its conjugate is also included in the model for the other final state with independent $c_j$ coefficients.  
States which can decay to $\KS\Kpm$, such as the $a_2(1320)^\pm$ particle, are also considered but none are found to contribute significantly.

The baseline model contains contributions from the $\Kstar(892)^{0,+}$, $K^*_0(1430)^{0,+}$ and $K^*_2(1430)^{0,+}$ resonances and their conjugates.
Thus ten parameters are determined from each Dalitz plot, corresponding to the magnitude and phase of the $c_j$ coefficient for each component except those for the $\KstarIzoptbar$ resonance which are fixed to serve as a reference amplitude.
The removal of any of these components from the model leads to deterioration of twice the negative log likelihood ($-$2NLL) by more than 25 units, while the addition of any other component does not improve $-$2NLL by more than 9 units.
The vector and tensor states are described with relativistic Breit--Wigner functions with parameters taken from Ref.~\cite{PDG2018}.
This is not appropriate for the broad $K\pi$ S-wave.
Several different lineshapes that have been suggested in the literature are tested, with the LASS description~\cite{lass} found to be most suitable in terms of fit stability and agreement with the data.  
The LASS shape combines the $K^*_0(1430)$ resonance with a slowly varying nonresonant component; the associated parameters are taken from Refs.~\cite{PDG2018,lass2}.
The combined shape is referred to as the $K^*_0(1430)$ component when discussing the amplitude fit; results for the resonant and nonresonant contributions are reported in addition to those for the total in Sec.~\ref{sec:results}.

The $\Bs\to \KstarIpm\Kmp$ and $\Bs \to \KstarIzoptbar\KorKbarz$ decays have previously been observed~\cite{LHCb-PAPER-2014-043,LHCb-PAPER-2015-018}.\footnote{
  The notation $\KstarIzoptbar\KorKbarz$ refers to the sum of the $\KstarIz\Kzb$ and $\KstarIzb\Kz$ final states.
}
The significance of each of the other contributions is evaluated using a likelihood ratio test.  
Ensembles of simulated pseudoexperiments are generated with parameters corresponding to the best fit to data obtained with models that do not contain the resonance of interest, but that otherwise contain the same resonances as the baseline model.
Each pseudoexperiment is fitted with models both including and not including the given resonance, from which a distribution of the difference in negative log likelihood is obtained.
This is found to be well fitted by a $\chisq$ shape, which can then be extrapolated to find the $p$-value corresponding to the $-$2NLL value obtained in data.  

Using this procedure, the significances for the \KstarIIp, \KstarIIz, \KstarIIIp\ and \KstarIIIz\ contributions are found to correspond to 17.3, 15.2, 4.0 and 4.8 standard deviations, when only statistical uncertainties are included.
The $K\pi$ S-wave contributions remain highly significant among all the systematic variations discussed in Sec.~\ref{sec:systematics}, and therefore the $\Bs\to \KstarIIpm\Kmp$ and $\Bs \to \KstarIIzoptbar\KorKbarz$ decays are observed with significance over 10 standard deviations.
However, some systematic variations do impact strongly on the need to include tensor resonances in the fit model, and thus preclude any similar conclusion for the $\Bs\to \KstarIIIpm\Kmp$ and $\Bs \to \KstarIIIzoptbar\KorKbarz$ decays.

\begin{figure}[!tb]
  \begin{center}
    \includegraphics*[width=0.47\textwidth]{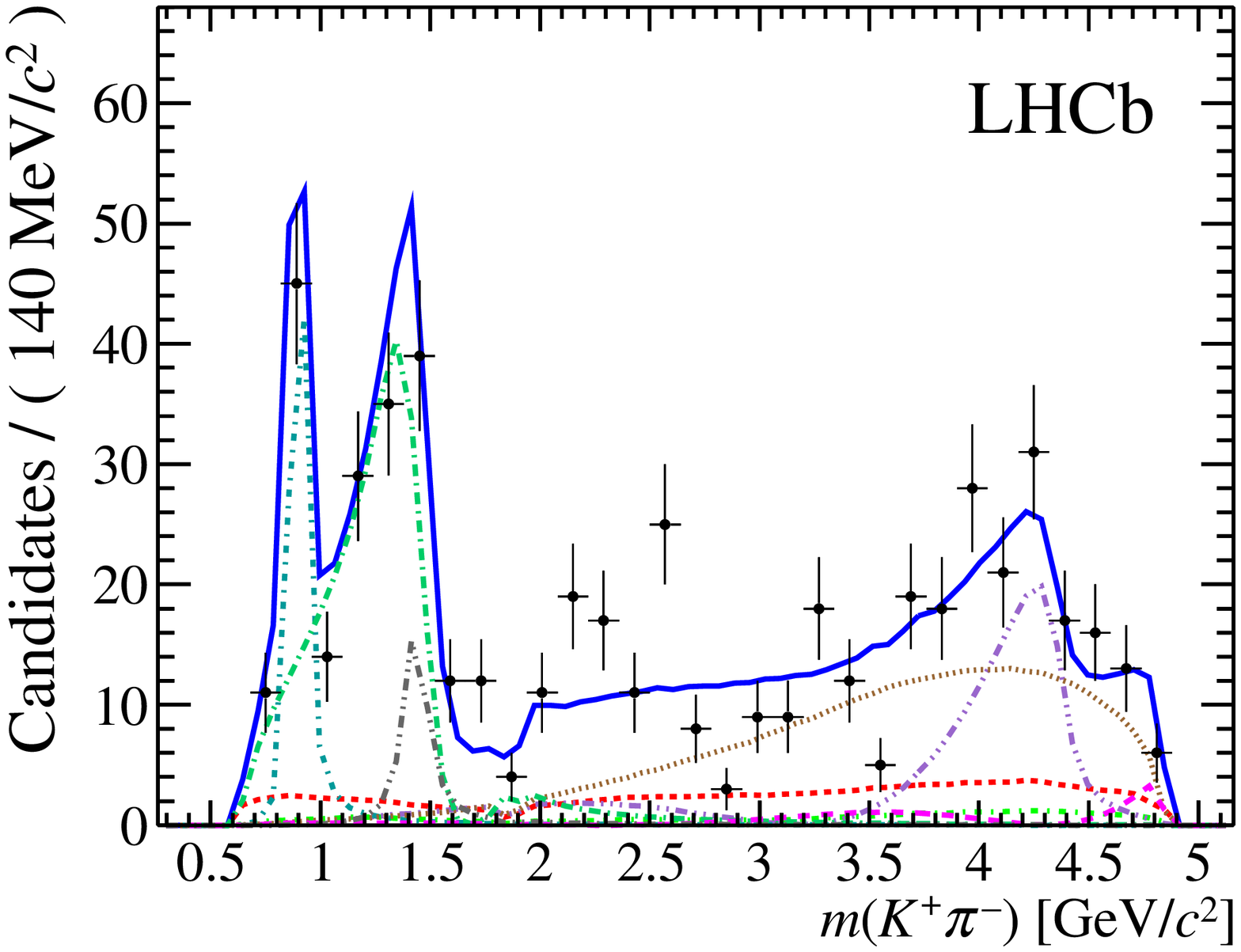}
    \includegraphics*[width=0.47\textwidth]{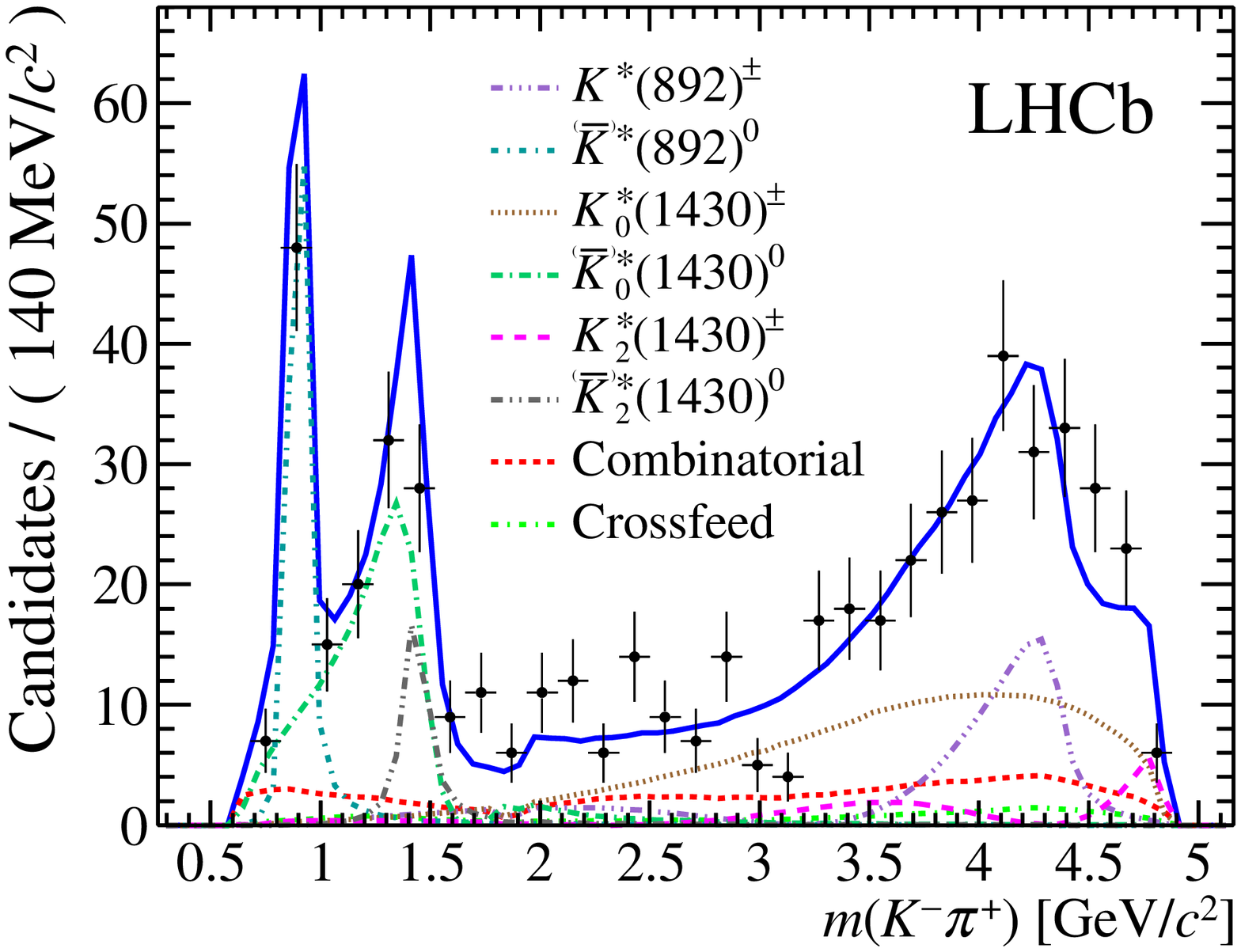}
    \includegraphics*[width=0.47\textwidth]{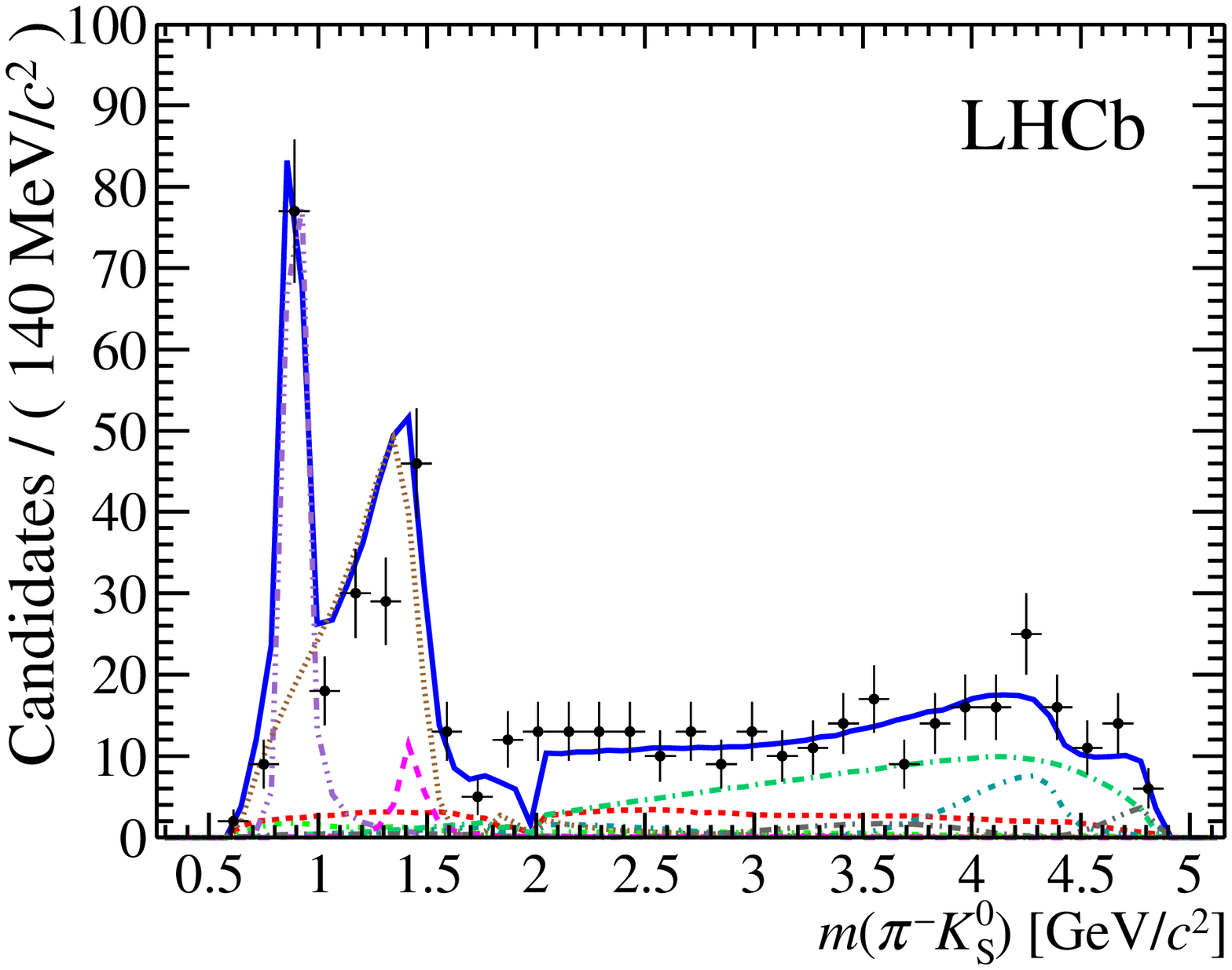}
    \includegraphics*[width=0.47\textwidth]{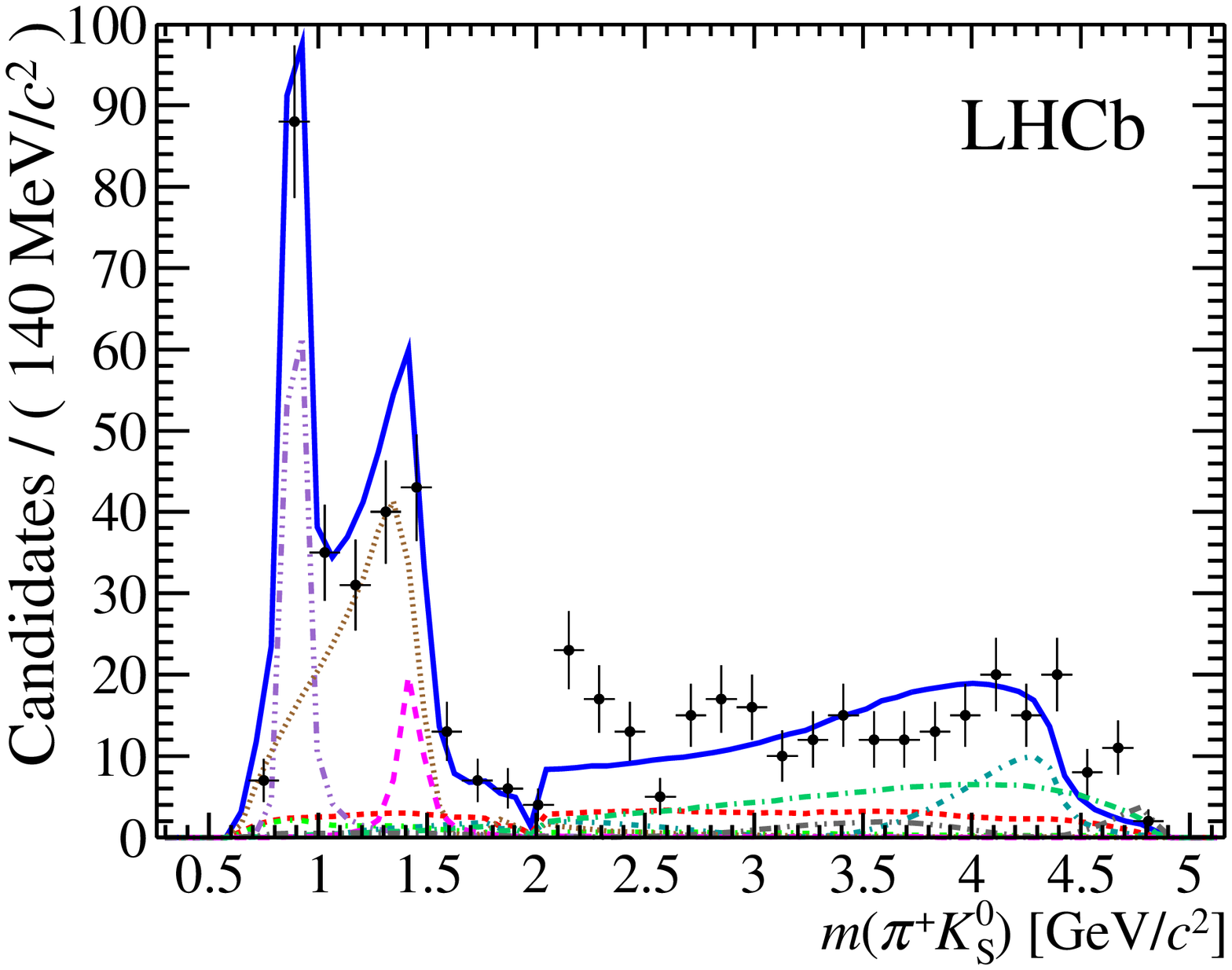}
    \includegraphics*[width=0.47\textwidth]{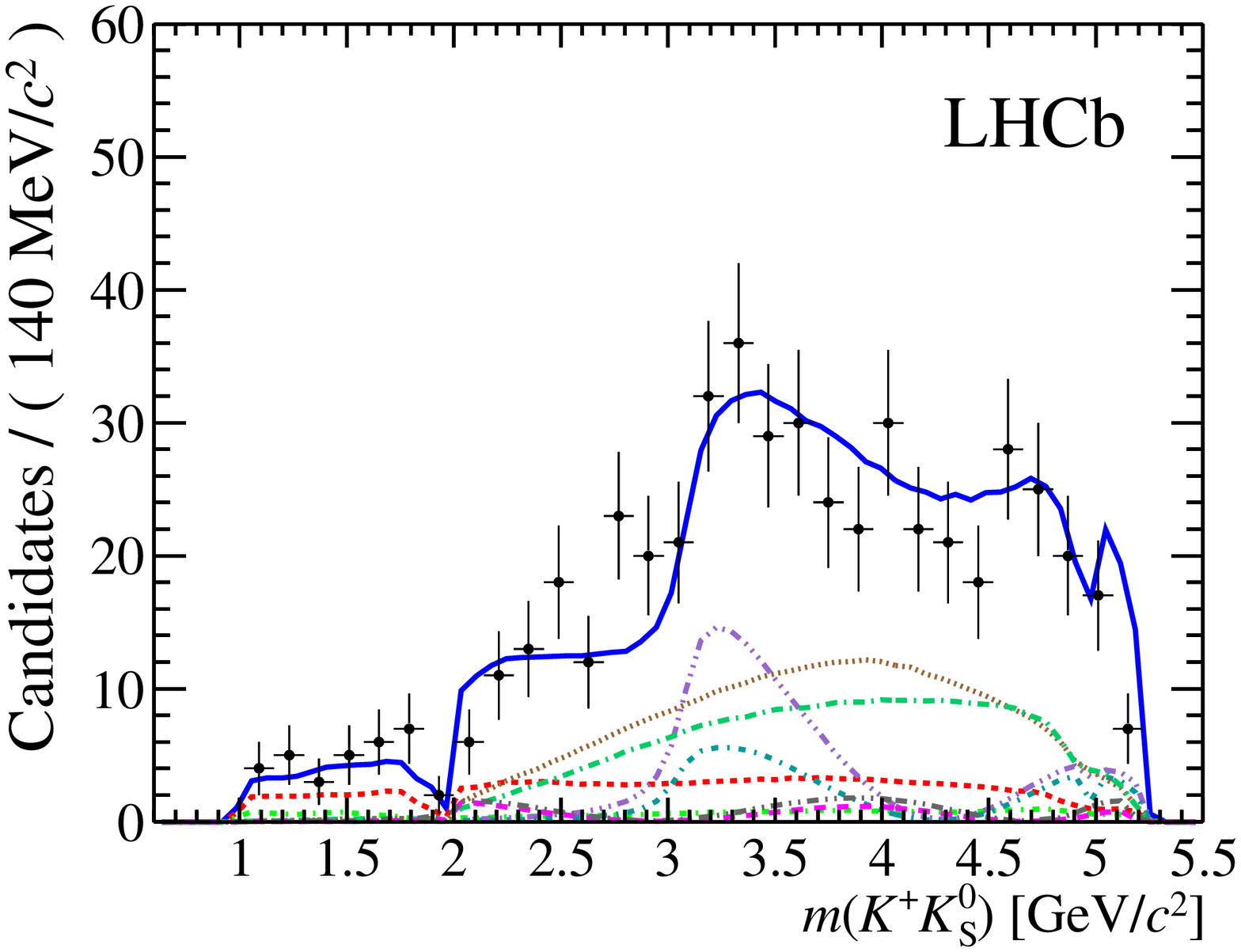}
    \includegraphics*[width=0.47\textwidth]{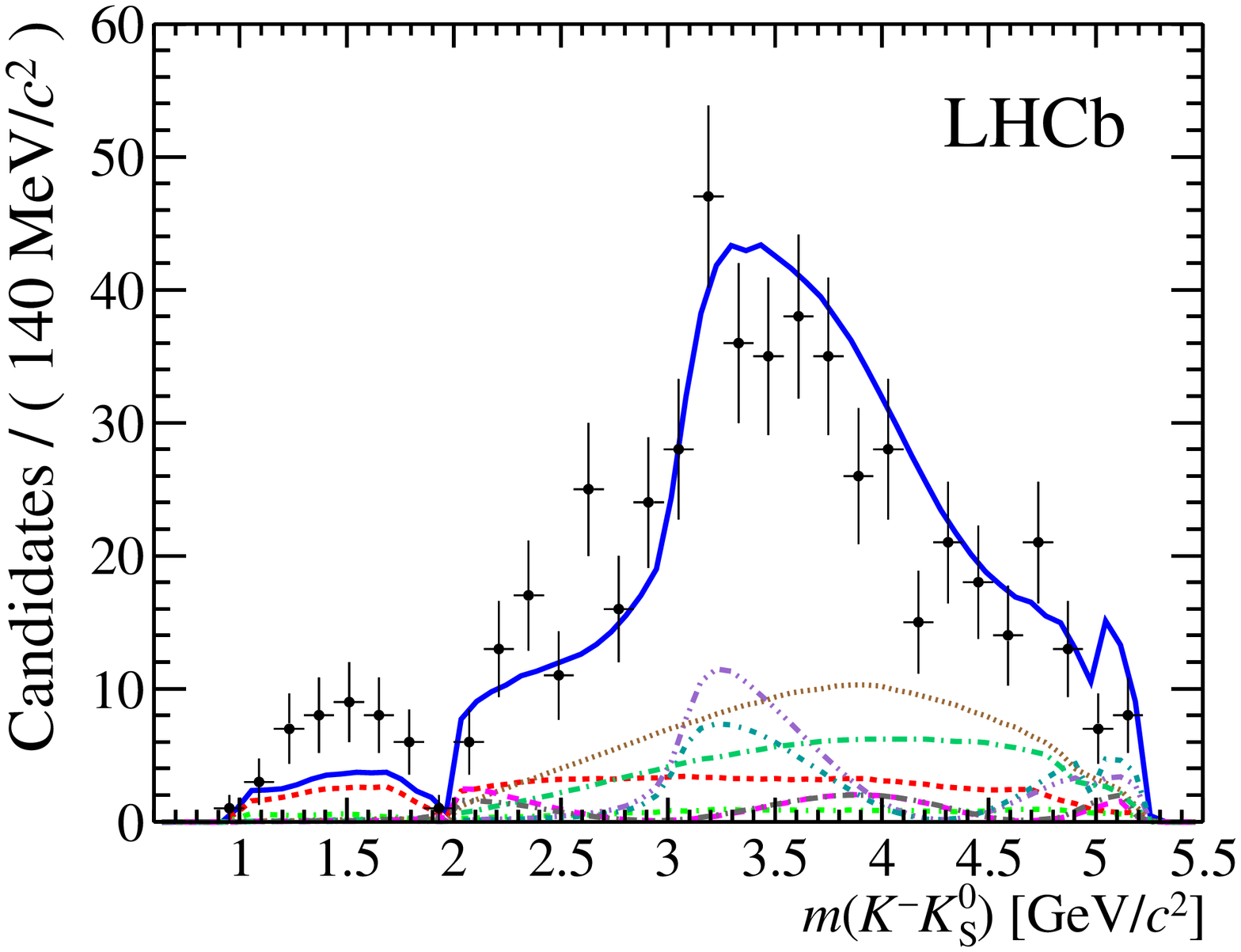}
  \end{center}
\caption{\small
  Invariant-mass distributions for (top)~$m(\Kpm\pimp)$, (middle)~$m(\KS\pimp)$ and (bottom)~$m(\KS\Kpm)$, for (left)~the $\KS\Kp\pim$ and (right)~the $\KS\Km\pip$ final states. 
  The data are shown with black points, while the full fit result is shown as a blue solid line with individual signal and background components as detailed in the legend.
  }
  \label{fig:dp-fits}
\end{figure}

The results of the fit of the baseline model to the data are shown in Fig.~\ref{fig:dp-fits}.
Various methods are used to assess the goodness-of-fit~\cite{Williams:2010vh} and good agreement between the model and the data is found. 
For example, using the point-to-point dissimilarity test, $p$-values of 0.27 and 0.19 are found for the \KsKpPim\ and \KsKmPip\ samples, respectively. 
The results for the fit fractions are given in Table~\ref{tab:FFs}.
The statistical uncertainties on the fit fractions are evaluated from the spreads in these values obtained when fitting ensembles of pseudoexperiments generated according to the baseline model with parameters corresponding to those obtained in the fit to data.
The fit fractions for each resonance and its conjugate (in the other Dalitz plot) are consistent, as expected from the absence of significant difference between the two Dalitz plot distributions seen in Figs.~\ref{fig:dp-distribution} and~\ref{fig:dp-fits}.
Thus, no significant \CP-violation effect is observed.

\begin{table}[!tb]
\centering
\caption{\small
  Results of the fit with the baseline model to the $\KS\Kp\pim$ and $\KS\Km\pip$ Dalitz plots.
  The fit fractions associated with each resonant component are given with statistical uncertainties only.
  The sums of fit fractions for both $\KS\Kp\pim$ and $\KS\Km\pip$ final states are 102\%, corresponding to low net interference effects.
  }
\label{tab:FFs}
\begin{tabular}{lclc}
\hline \\ [-2.4ex]   
\multicolumn{2}{c}{$\KS\Kp\pim$} & \multicolumn{2}{c}{$\KS\Km\pip$} \\
Resonance   & Fit fraction (\%)          & Resonance   & Fit fraction (\%)          \\
\hline \\ [-2.4ex]                                 
\KstarIm    & $15.6 \pm 1.5$             & \KstarIp    & $13.4 \pm 2.0$             \\
\KstarIIm   & $30.2 \pm 2.6$             & \KstarIIp   & $28.5 \pm 3.6$             \\
\KstarIIIm  & $\,\,\,2.9 \pm 1.3$        & \KstarIIIp  & $\,\,\,5.8 \pm 1.9$        \\
\KstarIz    & $13.2 \pm 2.4$             & \KstarIzb     & $19.2 \pm 2.3$             \\                                      
\KstarIIz   & $33.9 \pm 2.9$             & \KstarIIzb    & $27.0 \pm 4.1$             \\
\KstarIIIz  & $\,\,\,5.9 \pm 4.0$        & \KstarIIIzb   & $\,\,\,7.7 \pm 2.8$        \\
\hline
\end{tabular}
\end{table}

\section{Systematic uncertainties}
\label{sec:systematics}

Systematic uncertainties that affect the determination of the observables in the amplitude analysis arise from inaccuracy in the experimental inputs and the choice of the baseline amplitude parametrisation. 
The evaluation of effects arising from these sources is discussed in the following, with a summary of the systematic uncertainties on the fit fractions presented in Table~\ref{tab:sys-summary}. 

\begin{table}[tb]
\centering
\caption{\small
  Systematic uncertainties on the fit fractions, quoted as absolute uncertainties in percent. 
  The columns give the contributions from each of the different sources described in the text.
}
\label{tab:sys-summary}
\vspace{12pt}
\resizebox{\textwidth}{!}{
\begin{tabular}{lcccccccc|c}
\hline
& \multicolumn{9}{c}{Fit fraction (\%) uncertainties} \\
  Resonance   & Yields & Bkg. & Eff. & Fit bias & Add.\ res. & Fixed par. & Alt.\ model &  Method & Total  \\
\hline                                                                     
\KstarIm    & $0.2$   &   $0.2$   &   $0.5$  &   $0.2$   &   --      &   $0.7$    &   $\pho5.4$   &   $3.1$   &   $\pho6.3$ \\  
\KstarIIm   & $0.1$   &   $0.2$   &   $0.6$  &   $0.3$   &   $0.1$   &   $2.1$    &   $22.0$  &   $2.9$   &   $22.3$ \\
\KstarIIIm  & $0.1$   &   $0.1$   &   $0.3$  &   $0.6$   &   $0.1$   &   $1.8$    &   $\pho2.2$   &   $0.2$   &   $\pho2.9$ \\
\KstarIz    & $0.2$   &   $0.2$   &   $0.4$  &   $0.9$   &   --      &   $0.3$    &   $\pho7.0$   &   $2.0$   &   $\pho7.4$ \\
\KstarIIz   & $0.2$   &   $0.3$   &   $0.9$  &   $0.4$   &   $0.1$   &   $4.4$    &   $\pho3.3$   &   $1.3$   &   $\pho5.7$ \\
\KstarIIIz  & $0.1$   &   $0.3$   &   $0.7$  &   $1.3$   &   $0.2$   &   $4.4$    &   $\pho3.6$   &   $1.0$   &   $\pho6.0$ \\
\hline                                                                                                                
\KstarIp    & $0.4$   &   $0.1$   &   $0.6$  &   $0.5$   &   $0.1$   &   $0.7$    &   $\pho1.1$   &   $0.7$   &   $\pho1.8$ \\
\KstarIIp   & $0.5$   &   $0.4$   &   $0.7$  &   $0.8$   &   $0.2$   &   $6.4$    &   $13.0$  &   $4.5$   &   $15.2$ \\
\KstarIIIp  & $0.1$   &   $0.2$   &   $0.4$  &   $0.2$   &   $0.1$   &   $4.1$    &   $\pho4.5$   &   $3.2$   &   $\pho6.9$ \\
\KstarIzb   & $0.4$   &   $0.3$   &   $0.4$  &   $0.2$   &   $0.2$   &   $0.5$    &   $\pho3.0$   &   $7.9$   &   $\pho8.5$ \\
\KstarIIzb  & $0.4$   &   $0.4$   &   $0.6$  &   $0.8$   &   $0.7$   &   $0.9$    &   $\pho3.9$   &   $5.4$   &   $\pho6.8$ \\
\KstarIIIzb & $0.1$   &   $0.2$   &   $0.4$  &   $0.8$   &   $0.1$   &   $1.0$    &   $\pho5.5$   &   $2.7$   &   $\pho6.3$ \\
\hline        
\end{tabular}
}
\end{table}

Uncertainties associated with the signal and background yields obtained from the mass fit are examined by scaling the errors obtained from the whole mass fit range to the signal region. 
Statistical uncertainties on the yields are obtained from the covariance matrix of the baseline fit result, and systematic uncertainties are extracted similarly as for the branching fraction measurement~\cite{LHCb-PAPER-2013-042}.  
A series of pseudoexperiments are generated from the baseline mass fit, which are fitted by varying all of the fixed parameters according to their covariance matrix. 
The distributions of the differences from the baseline fit results are then fitted with a Gaussian function, and a systematic uncertainty is assigned as the sum in quadrature of the absolute value of the corresponding mean and width. 
The dependence on the models used in the invariant-mass fit is investigated by repeating the fit on ensembles generated with alternative shapes. 
The signal shape is examined by removing the tail to high mass values, whilst for the combinatorial background the effect of floating independently the slopes for each spectrum and replacing the exponential by a linear model are evaluated.  
These uncertainties are propagated into the amplitude fit by generating ensembles of pseudoexperiments in order to address the uncertainties related to the yield extraction, either by the RMS of the fitted quantity over the ensemble or the mean difference to the baseline model. 

Uncertainties arising from the modelling of the Dalitz plot distributions of both combinatorial and cross-feed backgrounds are estimated 
by varying the histograms used to describe these shapes within their statistical uncertainties in order to create an ensemble of new histograms.  
The data are refitted using each new histogram and the systematic uncertainty is taken from the RMS of the fitted quantity over the ensemble.

Effects related to the efficiency modelling are determined by repeating the Dalitz plot fit using new histograms obtained in a similar fashion as for the background. 
Uncertainties caused by residual disagreements between data and simulation are addressed by examining 
alternative efficiency maps, either by varying the binning-scheme choice or by using alternative corrections.
The simulated distributions of the features used in the BDT algorithm are known to have residual differences with respect to the data. 
The impact of this is estimated by repeating the amplitude fit using efficiency models that include additional corrections obtained with a multivariate weighting procedure~\cite{Rogozhnikov:2016bdp}.
Potential disagreements in the vertexing of the \KS meson as a function of momentum are also 
studied using $D^{*+}\to (D^{0}\to \phi \KS)\pi^{+}$ calibration samples, with a similar procedure to that used in Ref.~\cite{LHCb-PAPER-2012-009}.
Finally, effects related to the hardware-stage trigger are addressed
by calibrating the associated efficiency maps using $\Bp \to \jpsi \Kp$ and $\Bz \to \jpsi \Kp\pim$ control samples.  
The data fit is repeated including each of these new efficiency models and a systematic uncertainty is assigned from the mean difference to the results with the baseline model. 

Pseudoexperiments generated from the baseline fit results are used to quantify any intrinsic bias in the fit procedure. 
The uncertainties are evaluated as the sum in quadrature of the mean difference between the baseline and sampled values and the corresponding uncertainty. 

The choice of the baseline Dalitz fit model introduces important uncertainties through the choices of both the resonant or nonresonant contributions included and the lineshapes used.
The effects on the results of including additional $K^{*}(1410)$, $K^{*}(1680)$ or $a_{2}(1320)^{\pm}$ signal components in the fit are examined individually for each contribution. 
Some alternative fits give unrealistic results (for example, with very large sums of fit fractions) and are not included in the evaluation of this uncertainty.

Each resonant contribution has fixed parameters in the fit, which are varied to evaluate the associated systematic uncertainties. 
These include masses and widths~\cite{PDG2018} and the effective range and scattering length parameters of the LASS lineshape~\cite{PDG2018,lass2}.
The Blatt--Weisskopf radius parameter is varied within the range $3.0$--$5.0 \gev^{-1}\hbar c$.
The fit is repeated many times varying each of these fixed parameters within its uncertainties. 
The RMS of the distribution of the change in each fitted parameter is taken as the systematic uncertainty.

The baseline LASS parametrisation for the $K\pi$ S-wave modelling is known to be an approximate form, 
and associated uncertainties are assigned by evaluating the impact of an alternative parametrisation. 
This component is replaced by the model suggested in Ref.~\cite{ElBennich:2009da}, using tabulated magnitudes and phases at various values of $m(K\pi)$ obtained from form factors. 
This is found to provide a good description of the data, 
with changes in the fit fractions for the \KstarII\ terms partially compensated for by changes in interference effects between them. 
Further theoretical work is required to have an accurate description of the S-wave term, therefore 
the differences between this alternative model and the baseline model are conservatively assigned as systematic uncertainties.

The method of modelling each of the $\Bs\to\KS\Kpm\pimp$ Dalitz plots with a single amplitude is an approximation, as discussed in Sec.~\ref{sec:Introduction}.
The systematic uncertainty associated with the method is evaluated by generating with a full decay-time-dependent model ensembles of pseudoexperiments with different parameters for the contributing amplitudes based on the expected branching fractions~\cite{Cheng:2014uga,Li:2014fla} and a range of different \CP-violation hypotheses.
The results obtained from the fit with the approximate model are compared to those expected with the full model, with results for the fit fractions found to be robust (in contrast to the results for relative phases between resonant contributions).
The systematic uncertainty is assigned as the bias found in the case that the model is generated with the theoretically preferred values for the parameters~\cite{Cheng:2014uga,Li:2014fla}.

\section{Results}
\label{sec:results}

The flavour-averaged fit fractions are converted into products of branching fractions using Eq.~\eqref{eq:Chap-FFhat} and 
${\cal B}(\BsToKzBarOptKpi) = (84.3 \pm 3.5 \pm 7.4 \pm 3.4)\times10^{-6}$~\cite{LHCb-PAPER-2017-010}, to obtain
\begin{eqnarray*}
\Br{\Bs \to \KstarIpm\Kmp; \KstarIpm \to \KorKbarz\pipm}               ~~=& \\
& \hspace{-12mm}(   12.4 \pm   0.8   \pm   0.5   \pm   \phz2.7   \pm   1.3)  \times 10^{-6} \,, \\  
\Br{\Bs \to \KpiSpm\Kmp}                                               ~~=& \\
& \hspace{-12mm}(   24.9 \pm   1.8   \pm   0.5   \pm      20.0   \pm   2.6)  \times 10^{-6} \,, \\
\Br{\Bs \to \KstarIIIpm\Kmp; \KstarIIIpm \to \KorKbarz\pipm}           ~~=& \\
& \hspace{-12mm}(\phz3.4 \pm   0.8   \pm   0.4   \pm   \phz5.4   \pm   0.4)  \times 10^{-6} \,, \\
\Br{\Bs \to \KstarIzoptbar\KorKbarz; \KstarIzoptbar \to \Kmp\pipm}     ~~=& \\
& \hspace{-12mm}(   13.2 \pm   1.9   \pm   0.8   \pm   \phz2.9   \pm   1.4)  \times 10^{-6} \,, \\
\Br{\Bs \to \KpiSz\KorKbarz}                                           ~~=& \\
& \hspace{-12mm}(   26.2 \pm   2.0   \pm   0.7   \pm   \phz7.3   \pm   2.8)  \times 10^{-6} \,, \\
\Br{\Bs \to \KstarIIIzoptbar\KorKbarz; \KstarIIIzoptbar \to \Kmp\pipm} ~~=& \\
& \hspace{-12mm}(\phz5.6 \pm   1.5   \pm   0.6   \pm   \phz7.0   \pm   0.6)  \times 10^{-6} \,,
\end{eqnarray*}
where the uncertainties are respectively statistical, systematic related to experimental and model uncertainties, and due to
the uncertainty on ${\cal{B}}(\BsToKzBarOptKpi)$.\footnote{The notation \KpiS\ indicates the total \kpi\ S-wave that is modelled by the LASS lineshape.}
The experimental systematic uncertainties are those listed in Table~\ref{tab:sys-summary} due to signal and background yields, background and efficiency descriptions and fit bias,
while the model uncertainties are those related to the choice of resonances included in the baseline model, fixed parameters in the amplitude description, alternative models and the approach of modelling the two Dalitz plots with a single amplitude.
All statistical and systematic uncertainties are evaluated directly for the flavour-averaged fit fractions, rather than by propagating the uncertainties on the results separated by final state, to ensure that correlations are properly taken into account. 

It is possible to use the composition of the LASS lineshape to obtain separately the fractions of the contributing parts.
Integrating separately the resonant part, the effective range part, and the coherent sum, for both the \KpiSz and the \KpiSpm components,
the \KstarIIpm or \KstarIIzoptbar resonances are found to account for 78\%, the effective range term 46\%, and destructive interference between the two terms is responsible for the excess 24\%.
The branching fractions of the two nonresonant parts are found to be 
\begin{eqnarray*}
\Br{\Bs \to \KpiNRpm\Kmp}     &=& (11.4   \pm   0.8   \pm   0.2   \pm   9.2   \pm   1.2   \pm   0.5)  \times 10^{-6} \,, \\
\Br{\Bs \to \KpiNRz\KorKbarz} &=& (12.1   \pm   0.9   \pm   0.3   \pm   3.3   \pm   1.3   \pm   0.5)  \times 10^{-6} \,,
\end{eqnarray*}
where the fifth uncertainty is related to the proportion of the \KpiS\ component due to the effective range part.
Similarly, the products of branching fractions for the \KstarII\ resonances are
\begin{equation*}
\begin{array}{rcl}      
\Br{\Bs \to \KstarIIpm\Kmp; \KstarIIpm \to \KorKbarz\pipm}           &=& \\
& \multicolumn{2}{l}{\hspace{-14mm}(19.4   \pm   1.4   \pm   0.4   \pm    15.6   \pm   2.0   \pm   0.3)  \times 10^{-6} \,,} \\
\Br{\Bs \to \KstarIIzoptbar\KorKbarz; \KstarIIzoptbar \to \Kmp\pipm} &=& \\
& \multicolumn{2}{l}{\hspace{-14mm}(20.5   \pm   1.6   \pm   0.6   \pm \phz5.7   \pm   2.2   \pm   0.3)  \times 10^{-6} \,.}
\end{array}  
\end{equation*}

Results for the various \Kstar resonances are further corrected by their branching fractions to \kpi to obtain the quasi-two-body branching fractions.
The branching fractions to \kpi are~\cite{PDG2018}: $\Br{\KstarI \to \kpi} = 100\%$, $\Br{\KstarII \to \kpi} = (93 \pm 10)\%$ and $\Br{\KstarIII \to \kpi} = (49.9 \pm 1.2)\%$.
In addition, the values of $\Br{K^{*} \to K \pi}$ are scaled by the corresponding squared Clebsch--Gordan coefficients, 
\ie\ $2/3$ for both $\KstarzorKstarzb \to \Kpm \pimp$ and $\Kstarpm \to \KorKbarz \pipm$. 
Thus, the branching fractions are
\begin{eqnarray*}      
\Br{\Bs\to \KstarIpm\Kmp}              &=& (18.6      \pm   1.2   \pm   0.8   \pm   \phz4.0   \pm   2.0)  \times 10^{-6} \,, \\
\Br{\Bs\to \KstarIIpm\Kmp}             &=& (31.3      \pm   2.3   \pm   0.7   \pm      25.1   \pm   3.3)  \times 10^{-6} \,, \\
\Br{\Bs\to \KstarIIIpm\Kmp}            &=& (10.3      \pm   2.5   \pm   1.1   \pm      16.3   \pm   1.1)  \times 10^{-6} \,, \\
\Br{\Bs \to \KstarIzoptbar\KorKbarz}   &=& (19.8      \pm   2.8   \pm   1.2   \pm   \phz4.4   \pm   2.1)  \times 10^{-6} \,, \\
\Br{\Bs \to \KstarIIzoptbar\KorKbarz}  &=& (33.0      \pm   2.5   \pm   0.9   \pm   \phz9.1   \pm   3.5)  \times 10^{-6} \,, \\
\Br{\Bs \to \KstarIIIzoptbar\KorKbarz} &=& (16.8      \pm   4.5   \pm   1.7   \pm      21.2   \pm   1.8)  \times 10^{-6} \,,
\end{eqnarray*}  
where the uncertainties are respectively statistical, systematic related to experimental and model uncertainties, and due to
the uncertainty on ${\cal{B}}(\BsToKzBarOptKpi)$, $\Br{\Kstar\to\kpi}$ and,
in the case of \KstarII, the uncertainty of the proportion of the \KpiS
component due to the \KstarII resonance.

\section{Summary}
\label{sec:summary}

The first amplitude analysis of $\Bs \to \KS\Kpm\pimp$ decays has been presented, using a $pp$ collision data sample corresponding to $3.0\invfb$ collected with the LHCb experiment. 
A good description of the data is obtained with a model containing contributions from both neutral and charged resonant states $\Kstar(892)$, $K^*_0(1430)$ and $K^*_2(1430)$. 
No significant \CP-violation effect is observed.
Measurements of the branching fractions of the previously observed decay modes
\mbox{$\Bs\to\KstarIpm\Kmp$} and $\Bs\to\KstarIzoptbar\KorKbarz$ are consistent with theoretical predictions~\cite{Cheng:2014uga,Li:2014fla,Li:2018qrm} and also consistent with, but larger than, the previous LHCb results~\cite{LHCb-PAPER-2014-043,LHCb-PAPER-2015-018}, which they supersede.
This is partly due to the larger \BstoKsKPi\ branching fraction determined in the updated analysis based on both 2011 and 2012 data~\cite{LHCb-PAPER-2017-010} compared to the previous determination~\cite{LHCb-PAPER-2013-042}. 
This amplitude analysis provides better separation of the \KstarI\ states from the other contributions in the Dalitz plot, in particular the S-wave, and more accurate estimation of the associated systematic uncertainties.
Contributions from $K^*_0(1430)$ states are observed for the first time with significance above $10$ standard deviations. 

Increases in the data sample size will allow the reduction of both statistical and systematic uncertainties on these results.
As substantially larger samples are anticipated following the upgrade of LHCb~\cite{LHCb-TDR-012,LHCb-PII-EoI}, it will be possible to extend the analysis to include flavour tagging and decay-time-dependence, and therefore to obtain sensitivity to test the SM through measurement of \CP-violation parameters in $\Bs \to \KS\Kpm\pimp$ decays.

\section*{Acknowledgements}
\noindent We express our gratitude to our colleagues in the CERN
accelerator departments for the excellent performance of the LHC. We
thank the technical and administrative staff at the LHCb
institutes.
We acknowledge support from CERN and from the national agencies:
CAPES, CNPq, FAPERJ and FINEP (Brazil); 
MOST and NSFC (China); 
CNRS/IN2P3 (France); 
BMBF, DFG and MPG (Germany); 
INFN (Italy); 
NWO (Netherlands); 
MNiSW and NCN (Poland); 
MEN/IFA (Romania); 
MSHE (Russia); 
MinECo (Spain); 
SNSF and SER (Switzerland); 
NASU (Ukraine); 
STFC (United Kingdom); 
NSF (USA).
We acknowledge the computing resources that are provided by CERN, IN2P3
(France), KIT and DESY (Germany), INFN (Italy), SURF (Netherlands),
PIC (Spain), GridPP (United Kingdom), RRCKI and Yandex
LLC (Russia), CSCS (Switzerland), IFIN-HH (Romania), CBPF (Brazil),
PL-GRID (Poland) and OSC (USA).
We are indebted to the communities behind the multiple open-source
software packages on which we depend.
Individual groups or members have received support from
AvH Foundation (Germany);
EPLANET, Marie Sk\l{}odowska-Curie Actions and ERC (European Union);
ANR, Labex P2IO and OCEVU, and R\'{e}gion Auvergne-Rh\^{o}ne-Alpes (France);
Key Research Program of Frontier Sciences of CAS, CAS PIFI, and the Thousand Talents Program (China);
RFBR, RSF and Yandex LLC (Russia);
GVA, XuntaGal and GENCAT (Spain);
the Royal Society
and the Leverhulme Trust (United Kingdom);
Laboratory Directed Research and Development program of LANL (USA).

\addcontentsline{toc}{section}{References}
\setboolean{inbibliography}{true}
\bibliographystyle{LHCb}
\bibliography{main}
\newpage
\centerline
{\large\bf LHCb Collaboration}
\begin
{flushleft}
\small
R.~Aaij$^{29}$,
C.~Abell{\'a}n~Beteta$^{46}$,
B.~Adeva$^{43}$,
M.~Adinolfi$^{50}$,
C.A.~Aidala$^{77}$,
Z.~Ajaltouni$^{7}$,
S.~Akar$^{61}$,
P.~Albicocco$^{20}$,
J.~Albrecht$^{12}$,
F.~Alessio$^{44}$
% ...

{\footnotesize \it

% ...
$ ^{7}$Universit{\'e} Clermont Auvergne, CNRS/IN2P3, LPC, Clermont-Ferrand, France\\
% ...
$ ^{12}$Fakult{\"a}t Physik, Technische Universit{\"a}t Dortmund, Dortmund, Germany\\
% ...
$ ^{20}$INFN Laboratori Nazionali di Frascati, Frascati, Italy\\
% ...
$ ^{29}$Nikhef National Institute for Subatomic Physics, Amsterdam, Netherlands\\
% ...
$ ^{43}$Instituto Galego de F{\'\i}sica de Altas Enerx{\'\i}as (IGFAE), Universidade de Santiago de Compostela, Santiago de Compostela, Spain\\
$ ^{44}$European Organization for Nuclear Research (CERN), Geneva, Switzerland\\
% ...
$ ^{46}$Physik-Institut, Universit{\"a}t Z{\"u}rich, Z{\"u}rich, Switzerland\\
% ...
$ ^{50}$H.H. Wills Physics Laboratory, University of Bristol, Bristol, United Kingdom\\
% ...
$ ^{61}$University of Cincinnati, Cincinnati, OH, United States\\
% ...
$ ^{63}$Syracuse University, Syracuse, NY, United States\\
% ...
$ ^{77}$University of Michigan, Ann Arbor, United States, associated to $^{63}$\\
% ...
}
\end{flushleft}

\end{document}